\documentclass[sigconf,nonacm]{acmart}

\setcopyright{none}

{}
{}
{}
{}
{}

\newcommand{\benchmark}{AuditBench\xspace}
\newcommand{\totalScenarios}{51\xspace}
\newcommand{\eg}{e.g.,\xspace}

\newcommand{\lab}{Lab\xspace}
\newcommand{\optc}{OpTC\xspace}
\newcommand{\edge}{edge\xspace}
\newcommand{\raw}{raw\xspace}

\newcommand{\lm}{lateral movement\xspace}
\newcommand{\gpt}{GPT-5\xspace}
\newcommand{\gptmini}{GPT-5 mini\xspace}
\newcommand{\geminiflash}{Gemini 2.5 Flash\xspace}
\newcommand{\geminipro}{Gemini 2.5 Pro\xspace}
\newcommand{\llamamav}{Llama 4 Maverick\xspace}

\newcommand{\promptone}{Prompt v1\xspace}
\newcommand{\prompttwo}{Prompt v2\xspace}

\usepackage{tikz}
\usepackage{amsmath}
\usepackage{latexsym}      
\usepackage{mathtools}	   
\usepackage{xspace}   
\usepackage{booktabs}
\usepackage{epsfig}
\usepackage{xcolor}   
\usepackage{tabularx} 
\usepackage{makecell} 
\usepackage{pifont}
\usepackage[font=small,labelfont=bf]{caption}
\usepackage{balance}
\usepackage{multirow}
\usepackage{graphicx}
\usepackage{subcaption}
\usepackage{enumitem}
\usepackage{comment}
\usepackage{algorithm}
\usepackage{algpseudocode}
\usepackage{listings}
\usepackage[utf8]{inputenc}
\usepackage{hyperref}
\usepackage{xurl}
\usepackage[table,xcdraw]{xcolor}
\usepackage{colortbl}
\usepackage[table,dvipsnames]{xcolor}
\usepackage{float}
\usepackage{hyphenat}
\usepackage{soul}
\usepackage{cleveref}
\usepackage[most]{tcolorbox}

\definecolor{bestcellcolor}{gray}{0.9}
\definecolor{darkgreen}{HTML}{008000}

\begin{document}

\title{Benchmarking and Exploring the Capabilities of LLMs for Attack Investigations}

\author{Aniket Anand}
\affiliation{%
  \institution{University of Chicago}
  \country{USA}
}

\author{Yiwei Hou}
\affiliation{%
  \institution{University of California, Berkeley}
  \country{USA}
}

\author{Daniel Fields}
\affiliation{%
  \institution{University of Chicago}
  \country{USA}
}

\author{Alex Kantchelian}
\affiliation{%
  \institution{Google}
  \country{USA}
}

\author{David Tao}
\affiliation{%
  \institution{Google}
  \country{USA}
}

\author{Kurt Thomas}
\affiliation{%
  \institution{Google}
  \country{USA}
}

\author{Grant Ho}
\affiliation{%
  \institution{University of Chicago}
  \country{USA}
}



\renewcommand{\shortauthors}{Anand et al.}

\begin{abstract}

This paper presents \benchmark, a new benchmark dataset for evaluating the capabilities of LLMs at investigating security-related system audit logs. 
We design and use this benchmark to explore the performance of LLMs on four log-investigation tasks that incident response teams commonly perform, ranging from triaging alerts generated by detectors to identifying persistence mechanisms on compromised systems. 
\benchmark consists of system audit logs collected from Linux and Windows machines, and spans over 50 different security investigation scenarios, including both malicious and benign activity.
Using our benchmark, we evaluate and analyze the performance of five frontier LLMs at analyzing audit logs for attack investigations.
Our analysis illuminates how LLM performance and error profiles vary according to different design choices, such as differences in model size, data representation, prompt construction, and specific investigation tasks.
Additionally, we characterize the quality of the explanations produced by LLMs and the types of errors that models make across our benchmark. 
Collectively, our work provides a foundation for assessing the capabilities of LLMs for investigating security logs, novel insights for practitioners using LLMs in security operations, and important directions for future research.

\end{abstract}




\maketitle



\section{Introduction}

Many organizations employ dedicated security teams with expertise in analyzing network and system audit logs to detect, investigate, and recover from attacks.
Unfortunately, security teams suffer from an overwhelming volume of false alarms, resulting in slow remediation times and missed attacks~\cite{yangsocalerts,alahmadi2022-99fp,wang2025CAPTAIN}. 
Analyzing even a small number of alerts and their corresponding log data requires specialized knowledge and remains a challenging process. 
In particular, this forensic work requires analysts to not only reason about large amounts of data, but also bridge the semantic gap between low-level details reported by audit logs (e.g., system calls) and the higher-level activity that describes attackers' concrete behavior and outcomes. 

The emergence of large language models (LLMs) has led to significant interest and the development of many LLM-driven services for security forensics and log analysis~\cite{microsoft-security-pilot,crowdstrike-Charlotte,IBM-QRADAR}.
Despite this growing deployment and papers exploring various security applications of LLMs~\cite{zhang2025llms}, limited published work has explored the capabilities of LLMs for security audit log analysis and attack investigations.
This gap exists because of a lack of sufficiently diverse and well-labeled data for such evaluations, and the broad and amorphous nature of real-world attack investigations.

To address this problem, we develop and present a new evaluation benchmark, \textbf{\benchmark}, that assesses the capabilities of LLMs for analyzing audit logs relevant to security investigations.
Our benchmark spans four common security investigation tasks, with labeled log datasets spanning \totalScenarios distinct attack and benign scenarios.
We construct this dataset by combining both (1) newly-generated audit log data that captures a variety of benign behavior and dozens of simulated attack behaviors from the MITRE ATT\&CK framework~\cite{mitreattack}, and (2) carefully curated and labeled scenarios from the largest public dataset of system audit logs (DARPA OpTC~\cite{optc}).
As part of this benchmark, we develop an extensible and automated evaluation framework that computes a variety of performance metrics, ranging from true positive and false positive rates to the accuracy of LLM provided explanations. 

To demonstrate the utility of this benchmark, we conduct a detailed analysis of major LLMs to answer a range of interesting questions, such as: 
how well do LLMs perform at investigating security audit logs, and what is the quality of their explanations? 
Do different tasks or models exhibit performance trends? 
How do different prompt engineering decisions impact performance on these investigation tasks, and do prompt changes lead to general vs. model-dependent effects?
How does the data representation of audit log information impact the performance of LLMs?
To what extent do errors overlap across different models, and what are the characteristics of common classes of errors?
And, how might security teams and developers most effectively use LLMs for security log investigations?

Overall, we find that popular LLMs achieve mixed performance across the security investigation tasks in our benchmark, with a tendency towards overly-suspicious verdicts that produce many false positives (\S~\ref{sec:results_overall}). 
On one hand, several models exhibit strong performance (high true positive rates and low false positive rates) when tasked with identifying instances of data exfiltration.
However, across other tasks in our benchmark, LLMs frequently reported a wide spectrum of benign activity as instances of attacks.

Surprisingly, when comparing the performance of the largest LLMs versus smaller versions or open-weight models, our results show that larger models do not always achieve the best performance (\S~\ref{sec:results_size}). 
On multiple tasks, smaller models achieved equal or better performance (both higher true positive rates and lower false positive rates) than larger models.
Our work also finds that using a more compact preprocessed representation of audit logs leads to strictly better performance for some models across all tasks in our benchmark (\S~\ref{sec:results_representation}).
Although this transformed data representation doesn't improve performance for every model, our benchmark and corresponding analysis point to promising directions for practitioners and future work to reduce costs, while maintaining or improving LLM performance for security investigations.

When it comes to comparing and evaluating the performance of models against each other, our results suggest that prompt changes can have model-dependent effects (\S~\ref{sec:results_prompts}), similar to how the impact of data representation can vary across models.
Although some prompts led to strictly better performance for certain models across all tasks, 
a given set of prompt changes might cause improvements for some models while reducing the performance of others for the same task and data.
Taken together, we caution against using our benchmark, and potentially other benchmarks, as tools to compare models head-to-head without careful exploration and comparisons across the design space.
Rather, our experiments and analysis suggest that our benchmark can help developers and security teams study the effects of different design decisions for particular LLMs and systems in practice, and to assess whether any new design patterns enable improvement across a broad set of models.

Finally, our analysis of models' outputs suggests that LLM explanations for true instances of attack activity have fairly high accuracy (\S~\ref{sec:analysis}).
When they correctly identify attack behavior, LLM explanations tend to capture key attack behaviors reflected in our detailed groundtruth, without including spurious or incorrect details.
Nonetheless, our overall evaluation metrics suggest that models tend to flag many instances of benign activity as potential attacks.
Based on the LLMs' explanations for these false positives, it appears that models have high sensitivity to any anomalies in system object names and high volume activities (\S~\ref{sec:errors}).
Future work should explore ways to mitigate high volume of false positives and improve the robustness of LLMs' understanding of attack behavior.

Ultimately, our work provides an initial foundation for studying and improving the capabilities of LLMs for attack investigations.
First, we contribute a novel benchmark, AuditBench,\footnote{ \url{https://github.com/aanand300/auditlogsbench}} including all data, code, documentation, prompts, and model inputs/outputs from our experiments.
This benchmark enables both model developers to identify areas for improvement and practitioners to identify appropriate versus cautionary use cases and design patterns to leverage LLMs.
Second, we conduct an analysis using our benchmark along these lines, which uncovers opportunities for improving the cost and performance of LLMs for security investigations, as well as insights and advice for making design decisions and cross-model comparisons between LLMs.
\section{Background and Related Work}\label{sec:background}

In this work, we focus on the attack \textit{investigation} process.
Whereas detection involves generating alerts, security investigations consist of triaging alerts generated by existing detectors and analyzing different audit logs to understand the full scope of an attacker's activity across an organization's systems~\cite{socdef1,nistcsf,inamsokoakland,nist-incident-response}. 

\paragraph{System Audit Logs}
As part of an attack investigation, incident response guidelines recommend that security teams should analyze security-related audit logs to uncover key attack activities that require containment or remediation~\cite{nist-incident-response,cisa-incident-response,cmu-incident-response}.
This log data ranges from sources like network or endpoint intrusion detection systems to logs that operating systems capture via built-in capabilities, such as Linux auditd~\cite{auditd} and Windows ETW or Process Monitoring (procmon) logs~\cite{procmon}.
Our work focuses on this latter set of native host logs, specifically Linux auditd and Windows procmon, given their universal applicability and prevalence in academic work~\cite{inamsokoakland}.

These host logs capture low-level system call activity on a machine, such as read/write/execute operations performed by each process, and network connections initiated or received by each process.
Additionally, prior work has developed provenance systems, such as SPADE and CamFlow~\cite{spade,camflow}, that can restructure the information in these logs into a graph-based representation with edges that represent a subject, system call action, and an object receiving the action.
As discussed in Section~\ref{sec:data_representation}, our work also explores how different representations of log data impact an LLM's performance.

\paragraph{Automating Attack Investigations}
Prior work on automating attack investigation has studied using provenance graphs for detection and analysis.
These approaches include graph sketching~\cite{han2020unicorn} and various machine learning methods, ranging from anomalous node identification~\cite{wang2022threatrace} to GNNs~\cite{cheng2024kairos,rehman2024flash} and masked graph representation learning~\cite{jia2024magic} for improved detection.
Despite these advances, using provenance graphs for real-world investigations remains difficult due to the runtime overhead of these systems, as well as the difficulty and time it takes to interpret the attack graph outputs~\cite{dong2023we}.
Unlike potential LLM systems, these prior works require significant human interpretation and do not provide explainable verdicts in text form.

\paragraph{Using LLMs for Log Analysis}
Recent work has explored applying LLMs to log parsing, anomalous log event detection, and security-specific applications~\cite{akhtar2025llm-survey}.
For log parsing, systems such as LLMParser~\cite{ma2024llmparser} and LogParser-LLM~\cite{zhong2024logparser} show that LLMs can convert raw logs into structured formats with minimal rule engineering.
In anomaly detection applications, LogPrompt~\cite{liu2024logprompt} uses instruction-tuned prompts to detect abnormal log patterns and summarize failures without relying on training data.
Security-specific applications extend these ideas to threat detection:
LogBERT~\cite{guo2021logbert}, LogFiT~\cite{almodovar2022logfit}, and LogLLM~\cite{guan2024logllm} adapt transformer models for host-level anomalies, while APT-LLM~\cite{benabderrahmane2025aptllm}, Audit-LLM~\cite{song2024auditllm}, and RedChronos~\cite{li2025redchronos} apply LLMs to APTs, insider threat, and enterprise-scale log analysis.
Collectively, these studies demonstrate growing interest in using LLMs to automate log analysis and security investigation.
Beyond direct log analysis, Kramer et al.~\cite{kramer2025integrating} and Jones et al.~\cite{jones2025analysing} study LLM support for incident response and incident management, showing benefits for analyst assistance but also model-dependent limitations across response stages.
Our work advances this space by providing a new benchmark that allows the community to systematically study LLM-driven log analysis across diverse log representations and new security tasks and scenarios. 

\paragraph{Security Benchmarks for LLMs}
Existing security benchmarks for LLMs focus on three categories: CTF-style challenges~\cite{deng2024pentestgpt,shao2024nyuctfbench,zhang2025cybench}, vulnerability-focused evaluations (e.g., exploit generation and patching)~\cite{lee2025secbench,zhu2025cvebench,wang2025cybergym,zhang2025bountybench}, and cybersecurity knowledge tests assessing factual and procedural understanding~\cite{wang2025csebenchmark,vero2025baxbench}.
These benchmarks provide valuable insights into how LLMs reason about code and vulnerabilities, but different from our work, their tasks focus on offensive security and isolated software-level challenges, not security investigations.
A related line of work, LogEval~\cite{cui2025logeval}, benchmarks LLMs on general log analysis tasks such as parsing and anomaly detection in AIOps settings.
While conceptually close, LogEval focuses on system reliability, fault diagnosis, and similar benign kinds of anomalies, which are fundamentally different than security incidents and attacks.
Another work, ExCyTIn-Bench~\cite{wu2025excytin}, presents a benchmark consisting of logs from Microsoft-specific services such as Azure and MS Sentinel logs.
The benchmark constructs granular question-answer pairs for evaluating the performance of LLMs, and provides models with query access to a log database to answer questions about an incident.
Our benchmark expands upon this space by broadening the set of logs and security tasks under evaluation.
In doing so, we enable novel and complementary findings, as demonstrated by our empirical analysis (\eg the potential opportunities of pre-processing and transforming log data representations).
\begin{table}[t!]
\centering
\resizebox{\linewidth}{!} {%
\begin{tabular}{@{}lcrrrr@{}}
\toprule
\textbf{Task} & \textbf{Scenario} & \makecell[r]{\textbf{\# Lines} \\ \textbf{(Raw)}} & \makecell[r]{\textbf{\# Lines} \\ \textbf{(Edge)}} & \textbf{Duration} & \textbf{\# TP} \\
\midrule
\multirow{5}{*}{Classification}
& 1 & 1,229,728 & 5,198 & 300s & 1 \\
& 2 & 238 & 68 & 60s & 1 \\
& 3 & 8,843 & 1,493 & 240s & 1 \\
& 4 & 10,980 & 2,618 & 600s & 1 \\
& 5 & 369,109 & 2,064 & 360s & 1 \\
\addlinespace
\multirow{7}{*}{\begin{tabular}[c]{@{}l@{}}Benign\end{tabular}} 
& 1 & 6,954 & 1,857 & 180s & 0 \\
& 2 & 9,394 & 2,120 & 300s & 0 \\
& 3 & 1,399 & 509 & 300s & 0 \\
& 4 & 6,616 & 1,443 & 300s & 0 \\
& 5 & 5,869 & 1,207 & 720s & 0 \\
\addlinespace
\multirow{5}{*}{\begin{tabular}[c]{@{}l@{}}Lateral\\Movement\end{tabular}} 
& 1 & 1,782,760 & 3,189 & 180s & 1 \\
& 2 & 1,126,082 & 6,569 & 360s & 1 \\
& 3 & 278,413 & 1,051 & 240s & 1 \\
& 4 & 333,187 & 1,978 & 120s & 1 \\
& 5 & 193 & 27 & 177s & 1 \\
\addlinespace
\multirow{5}{*}{Persistence} 
& 1 & 495 & 140 & 300s & 1 \\
& 2 & 833 & 188 & 480s & 1 \\
& 3 & 642 & 304 & 360s & 1 \\
& 4 & 1,538 & 437 & 600s & 1 \\
& 5 & 771 & 207 & 300s & 1 \\
\addlinespace
\multirow{5}{*}{\begin{tabular}[c]{@{}l@{}}Data\\Exfiltration\end{tabular}} 
& 1 & 58 & 13 & 120s & 1 \\
& 2 & 2,156 & 947 & 2,160s & 1 \\
& 3 & 15,490 & 3,801 & 1,560s & 1 \\
& 4 & 2,331 & 634 & 300s & 1 \\
& 5 & 722,782 & 3,917 & 300s & 1 \\
\bottomrule
\end{tabular}
}
\caption{The Lab dataset portion of our benchmark consists of 25 unique scenarios, with 5 scenarios per security investigation task and 5 scenarios of purely benign activity (\S~\ref{sec:benchmark:data:lab}).}
\label{tab:dataset-lab-overview}
\end{table}

\begin{table}[ht]
\centering
\resizebox{\linewidth}{!} {%
\begin{tabular}{@{}lcrrr@{}}
\toprule
\textbf{Task} & \textbf{Scenario} & \makecell[r]{\textbf{\# Lines} \\ \textbf{(Edge)}} & \textbf{Duration} & \textbf{\# TP} \\
\midrule
\multirow{5}{*}{Classification}
& 1 & 1,077 & 10s & 1 \\
& 2 & 1,772 & 4s & 1 \\
& 3 & 1,833 & 21s & 1 \\
& 4 & 2,714 & 118s & 1 \\
& 5 & 2,248 & 93s & 1 \\

\addlinespace
\multirow{5}{*}{Benign} 
& 1 & 10,573 & 60s & 0 \\
& 2 & 4,543 & 60s & 0 \\
& 3 & 5,198 & 60s & 0 \\
& 4 & 2,784 & 60s & 0 \\
& 5 & 10,950 & 60s & 0 \\

\addlinespace
\multirow{5}{*}{Velox Benign} 
& 1 & 7,000 & 20s & 0 \\
& 2 & 1,106 & 20s & 0 \\
& 3 & 7,931 & 20s & 0 \\
& 4 & 5,281 & 20s & 0 \\
& 5 & 1,033 & 20s & 0 \\

\addlinespace
\multirow{5}{*}{\begin{tabular}[c]{@{}l@{}}Lateral\\Movement\end{tabular}} 
& 1 & 355 & 20s & 3 \\
& 2 & 910 & 20s & 9 \\
& 3 & 2,041 & 20s & 2 \\
& 4 & 463 & 20s & 2 \\
& 5 & 953 & 20s & 5 \\
\addlinespace
\multirow{4}{*}{Persistence} 
& 1 & 262 & 20s & 1 \\
& 2 & 2,519 & 20s & 1 \\
& 3 & 3,845 & 20s & 1 \\
& 4 & 1,209 & 20s & 1 \\
\addlinespace
\multirow{2}{*}{\begin{tabular}[c]{@{}l@{}}Data\\Exfiltration\end{tabular}} 
& 1 & 606 & 20s & 1 \\
& 2 & 2,943 & 60s & 1 \\
\bottomrule
\end{tabular}
}
\caption{The OpTC dataset portion of our benchmark consists of 26 scenarios curated and labeled from the DARPA OpTC dataset, including 10 scenarios of benign activity. For each benchmark investigation task, we identified either five scenarios or as many as we could based on the official DARPA groundtruth (\S~\ref{sec:benchmark:data:optc}).}
\label{tab:dataset-optc-overview}
\end{table}

\section{Benchmark Dataset}\label{sec:benchmark}
Our benchmark dataset covers 51 scenarios across four security investigation tasks (\S~\ref{sec:benchmark:tasks}).
Since our benchmark aims to evaluate performance on \textit{investigation} tasks, our scenarios presume that an analyst and LLM already have a concrete alert or pointer to anchor their analysis around a \textit{targeted set of log lines}~\cite{nistcsf,inamsokoakland}.
Tables~\ref{tab:dataset-lab-overview} and~\ref{tab:dataset-optc-overview} summarize the size and statistics of our benchmark dataset.

The data consists of operating system logs spanning diverse attack and benign behavior, along with detailed and manually-verified ground truth labels.
We generated 25 scenarios on virtual machines (\textbf{\lab data}: \S~\ref{sec:benchmark:data:lab}), and curated and labeled an additional 26 scenarios from the public DARPA OpTC dataset~\cite{optc} (\textbf{\optc data}: \S~\ref{sec:benchmark:data:optc}). 
For space constraints, we provide a high-level description of our datasets and labeling procedures, with additional details in Appendix~\ref{sec:appendix:dataset:optc}.
We publicly release our paper’s data and code, including ground-truth labels and log-capturing configurations. 

\subsection{Security Investigation Tasks}\label{sec:benchmark:tasks}
Our framework consists of 4 attack investigation tasks.
These tasks encompass an alert triaging task and three detailed investigation tasks that security teams perform after discovering an incident.
We selected these 4 tasks from recommended actions that consistently appeared in both government incident response frameworks, from NIST~\cite{nelson2025incident} and CISA~\cite{cisaplaybook}, as well as industry guides for threat-hunting and investigation (such as MITRE's threat hunting guide~\cite{daszczyszak2019ttp}). 

\begin{enumerate}
    \item Classification (Alert Triaging): given an alert and relevant audit log data, determine whether the alert corresponds to a real attack or a false alarm.
    \item Persistence investigation: identify and explain activity that creates persistence mechanisms in a set of log data (that could allow an attacker to maintain access to the system, as described by MITRE ATT\&CK Tactic TA0003~\cite{mitreattack}). 
    \item Lateral movement investigation: identify and explain all potential lateral movement actions in a set of log data (outbound connections that could provide remote access and control for an attacker; MITRE ATT\&CK Tactic TA0008). 
    \item Data exfiltration investigation: identify and explain potential events where sensitive user data is exfiltrated from the system in a set of log data (MITRE ATT\&CK Tactic TA0010). 
\end{enumerate}

\paragraph{Groundtruth Labels}\label{sec:ground_truth_labels}
For every scenario, our evaluation framework asks models to output information such as the activity timestamps, key log lines, and explanations for a model's decision.
We also developed task-specific groundtruth labels for automatic evaluations that check the following key fields in a model's output.
\begin{enumerate}
    \item Classification: a binary value of whether the scenario is malicious or benign.
    \item Persistence: (1) the MITRE ATT\&CK name of the persistence technique and (2) the timestamp(s) of where persistence mechanisms were created.
    \item Lateral Movement: (1) the timestamp of an outbound connection with remote execution capabilities on the target device and (2) the host identifier of the target. 
    \item Exfiltration: (1) the data (file path and file name) exfiltrated to an external device and (2) timestamp of the data movement.
\end{enumerate}

\subsection{Data: \lab data}\label{sec:benchmark:data:lab}

Although we identified a set of viable scenarios from some public datasets (\S~\ref{sec:benchmark:data:optc}), many existing datasets such as DARPA TC3~\cite{darpa-tc} lack sufficient coverage of attack events across our four tasks and/or lack sufficiently precise ground truth labels of the logs.
As a result, we created a new audit log dataset of attacks and benign activity on Linux and Windows machines. 
We constructed 5 distinct attack scenarios for every investigation task to cover diverse real-world attack behavior. Each scenario captures a unique attack technique (or sequence of techniques for some \textit{attack classification} task) based on the MITRE ATT\&CK framework~\cite{mitreattack}. 
Additionally, we generated five benign scenarios consisting of different, purely benign activities.
In total, this \lab dataset consists of audit log data for 25 unique scenarios. 
As discussed in Section~\ref{sec:data_representation}, scenarios in our \lab data have two different data representations: (1) a \textbf{\raw} representation (data directly from the native OS log collector), (2) an \textbf{\edge} representation (after pre-processing). 


\paragraph{Scenario Design}
To execute the MITRE ATT\&CK techniques for a malicious scenario, we studied and used both MITRE ATT\&CK's documentation and the Atomic Red Team (ART) framework~\cite{atomic}.

The 3 detailed investigation tasks of \benchmark, lateral movement, persistence and exfiltration, each correspond to a high-level tactic in MITRE ATT\&CK, and have a list of ``techniques" (concrete methods) for executing that tactic. 
For each investigation task (ATT\&CK tactic), we selected 5 distinct techniques that were feasible on Linux or Windows VMs and would generate observable activity in audit logs. 
For the classification task, we reviewed more techniques across the full set of ATT\&CK tactics, and selected ones that clearly indicated compromise of the system so that we could have accurate and automatic binary labels for this task's groundtruth.
Four of our classification task scenarios consist of a single ATT\&CK technique, whereas the final one consists of a combination of multiple ATT\&CK techniques. 
After identifying the techniques for each scenario, we followed the attack simulation steps from ART~\cite{atomic} for these techniques which occasionally include additional steps (such as deletion of keys after encrypting sensitive files, etc.) in some scenarios to ensure the scenarios were clearly malicious.

Finally, for our five benign scenarios, we simulated different legitimate user operations, such as system administration tasks, normal software installations and programming, file management, and web browsing/network communications.
These scenarios cover a mix of GUI-based applications and command-line tools.
Appendix~\ref{sec_appendix_dataset_lab} contains additional specific details of our scenarios. The start and end of each scenario corresponds to the start and end of the deliberate benign user activity that was performed.


\paragraph{Data collection} 
For each scenario, we first documented the list of steps for simulating it (\eg the applications and commands to execute).
Then we simulated these steps using VMs with one of the Windows 10, Windows Server 2022 or Ubuntu 20.04 LTS based on ART's specifications.
For log collection, we used ProcMon~\cite{procmon} on Windows machines and Auditd~\cite{auditd} on Linux machines. 

After setting up the log collection environment, we simulated attack and benign events based on the steps documented for every scenario. We recorded the timestamps of execution of the steps in our documentation for building the ground truths(\S~\ref{sec:ground_truth_labels}).
No other explicit user activity was performed during the execution of the benign or attack steps. 
We include this documentation as part of our open-sourced data.


\subsection{Data: \optc data}\label{sec:benchmark:data:optc}
Our benchmark also contains 26 scenarios curated from the DARPA OpTC~\cite{optc}, the largest publicly available system audit log dataset. 
Compared to other datasets, \optc contains the most diverse variety of \textit{attack} activity, including lateral movement, persistence, and exfiltration events, and the largest amount of benign activity. 
The \optc data creators published a ground truth PDF with a timeline and description of attacks, enabling us to construct attack scenarios and with accurate ground truth labels(\S~\ref{sec:ground_truth_labels}). 
Despite its size, we could identify 4 persistence scenarios and 2 exfiltration scenarios based on a detailed examination of the official ground truth by three researchers.
Unlike our \lab data, DARPA OpTC's data comes in only a pre-processed \edge representation and does not have the logs in their raw, native OS form.


\paragraph{Attack Scenario Design}
To construct our 16 attack scenarios, we initially attempted to directly use the groundtruth to find log lines for the timestamps and hosts with attack activity matching our tasks.
Unfortunately, the timestamps of documented malicious activity in the OpTC groundtruth often differed by seconds from when the activity occurred in the log data.
As a result we followed a multi-step process with manual verification for constructing attack scenarios from this dataset.
First, three researchers reviewed the official groundtruth to identify instances of attack activity that mapped to the four tasks in our benchmark, based on the MITRE ATT\&CK tactic definition or where operations clearly indicate malicious behavior that attempts to compromise the system security for the classification task (e.g. credential theft attempts by running known malicious software such as Mimikatz).
Next, for each instance that mapped to a potential investigation scenario, the researchers documented as many of the following key identifiers in the groundtruth document for the scenario as we could find (a subset of 8 identifiers: timestamps, file or registry name, PID, process name, IP address, port number, hostname, and principal/username).
With this set of identifiers, this set of researchers searched the audit log data for matches close to the reported timestamps to find matching attack ``event'' log lines.
We then constructed each scenario by taking a 20 second window of log line data surrounding the attack event lines (10s before and after).

For some classification and exfiltration tasks, we needed to select longer time durations to ensure that the log lines contained the full activity; for two classification tasks, we had shorter duration scenarios since we found the target attack action in just a single second.
Additionally, as shown in Table~\ref{tab:dataset-optc-overview}, some investigation scenarios have multiple instances of the attack activity during a scenario's time window (e.g., multiple instances of lateral movement or exfiltration).\footnote{We verified all attack activity instances in each scenario by having two researchers create an information flow graph from the scenario's data, and then analyze whether any task-specific attack patterns existed (e.g., for lateral movement, whether an edge had a process make an outbound network connection to an external machine, where the process had remote execution capabilities). 
Our released dataset also contains documentation describing our detailed groundtruth labeling process.}

\paragraph{Benign Scenario Design}
Our benchmark contains 10 benign scenarios curated from the \optc dataset.
To construct our first five OpTC benign scenarios, we selected 5 different hosts with no documented attack involvement or activity in the groundtruth, and then selected logs from 60-second windows at a random time on each host.
Next, we created an additional five OpTC benign scenarios specifically designed to test an LLM's capabilities to triage false positive detection alerts.
For these additional scenarios, we ran a recent state-of-the-art PIDS detector (Velox~\cite{bilot2025sometimes}) on log data from three hosts during a timespan that preceded the first day of any attack activity.
Based on this setup, we identified five false positive alerts by selecting five nodes flagged by Velox using its code's preset threshold. 
We then constructed five benign scenarios by identifying the first occurrence of each of these false positive nodes on the relevant host and taking all log lines in a 20-second window around it (10s before and after).

\subsection{Data Representation }\label{sec:data_representation}

As part of our benchmark, we explore two different \textit{representations} of audit log data: a \textit{raw} log and an \textit{edge} representations.
The raw representation is the exact log output recorded by native logging tools like Linux's \texttt{auditd}.
The edge representation is a processed version of this raw data that first converts the raw logs into a provenance or information flow graph. 
Each line in the edge representation corresponds to an edge in the information flow graph, with a subject node, object node, and action; and data in the edge representation is sorted by timestamp similar to the raw logs.
We use a standard provenance tool (SPADE~\cite{spade}) to generate an edge representation of the raw logs for the \lab portion of our benchmark data.

Exploring both data representations allows for us to have normalized audit log representations across different operating systems (e.g., Linux vs. Windows), and across different datasets (e.g., DARPA OpTC only provides logs in MITRE CAR formats~\cite{car,ecar}, which can easily be transformed into an edge representation but not back to raw logs). 
Furthermore, the raw Windows procmon logs are simply too verbose and voluminous to be practically used as inputs to LLMs (Table~\ref{tab:dataset-lab-overview}), with millions of log lines generated in just a few minutes.
In contrast, the edge representation greatly reduces redundant log events and therefore allows us to explore tradeoffs of using compact representations of log data.
For our experiments with our \lab data (\S~\ref{sec:results}), we only run evaluations with the raw logs for scenarios performed on Linux machines; 
a total of 8 scenarios for our classification task (including benign scenarios), 6 for lateral movement, 10 scenarios for persistence, and 9 for exfiltration.
\section{Evaluation Pipeline}\label{sec:methodology}
Our benchmark takes an LLM (\S~\ref{sec:models_and_parameters}), a scenario's log data (\S~\ref{sec:benchmark} and \S~\ref{sec:evaluation_inputs}), a task-specific prompt (\S~\ref{sec:prompt_construction}), and 
computes a set of performance metrics by comparing the LLM's output against our groundtruth labels (\S~\ref{sec:evaluation_metrics}).

\subsection{Models and Parameters}\label{sec:models_and_parameters}
We used our benchmark to evaluate 5 frontier LLMs, including two of the largest state-of-the-art models (\geminipro and \gpt) and smaller frontier models (\geminiflash, \gptmini, and \llamamav).
To ensure deterministic and minimized variability in responses, we used a temperature of 0 for all experiments when available.\footnote{\gpt does not provide temperature values other than the default value of 1.}
For experiments with the GPT-5 models, we set the reasoning effort to the ``medium'' default.

\subsection{Model Inputs}\label{sec:evaluation_inputs}
To evaluate a model's performance on an individual scenario, our benchmark runs the model with a common prompt for the scenario's investigation task with the scenario's audit log data.
Each prompt specifies a structured JSON output format, which our pipeline automatically compares against our groundtruth labels (\S~\ref{sec:ground_truth_labels}).

\subsubsection{Log Input}\label{sec:evaluation_chunking}
In several of our scenarios, and in many real-world settings, the audit log data might exceed the context window (token limit) for an LLM.
As a result, our evaluation pipeline breaks a scenario's audit log data into fixed-sized \textit{chunks} of sequential log lines that (including the task prompt) will fit within context windows of 128,000 tokens (a common minimum window size among many models).
When evaluating scenarios on their raw log representation, we used chunks of 1,000 lines; and for evaluations on the edge representation, we used chunks of 400 lines, where these chunk sizes come out to 100,000 - 128,000 tokens based on our data.
As described later in Section~\ref{par:chunk_size}, we also evaluated the impact of different chunk sizes for larger-context models.

Our benchmark evaluates a scenario by iteratively evaluating the performance on each chunk (plus the task prompt).
It then concatenates the outputs from each chunk into a global set for the scenario and computes final evaluation metrics over this global output set  (\S~\ref{sec:evaluation_metrics}).

\subsubsection{Prompts}\label{sec:prompt_construction}
For each of the four tasks, our benchmark uses a common prompt to evaluate a model's performance across all the scenarios in a task.
Each task's prompt has minor changes for how it describes the audit log data format when evaluating on raw vs. edge representations.
In our benchmark and experiments, we studied best practices from several leading LLM platforms~\cite{GoogleGeminiPrompting},~\cite{AzureAIFoundryPromptEngineering},~\cite{AnthropicContextEngineering} and constructed two versions of each task's prompt.
In our public code/data, we include both prompt versions and copies of the inputs and outputs for all models across our experiments (\S~\ref{sec:results}).
Future work can easily extend the benchmark or evaluate models using different prompts.

\paragraph{Prompt Version 1}
Our first prompt followed a general template with five components:
(1) task goals and definitions (e.g., definitions of persistence from the MITRE ATT\&CK framework),
(2) output format specifications that stated the key fields and what to output,
(3) a version of chain-of-thought that required the model to output explanations for the model's reasoning,
(4) descriptions of the input data and format,
(5) instructions to analyze all of the provided input data and log lines.
Four researchers drafted a prompt for each task, then met together to iteratively refine and finalize the prompt (e.g., agreeing on the language to reduce ambiguities and ensure output requirements matched our groundtruth formatting).
Appendix~\ref{sec:appendix:prompt} provides the general template and an example of the prompt used for the exfiltration task.

\paragraph{Prompt Version 2}
We also designed a second version of each task's prompt, as a way to explore the impact of prompt engineering and to simulate a practical setting where an analyst refined their prompts based on their initial usage.
Specifically, two researchers (including one not involved in the first prompt version creation) reviewed the latest prompt engineering best practices~\cite{GoogleGeminiPrompting, AzureAIFoundryPromptEngineering, AnthropicContextEngineering} and edited the Version 1 prompts in four ways.
First, these researchers interactively refined each task's prompt by looking at a set of false positive errors of two scenarios with higher FPs in each task from our \lab dataset.
Second, we increased the specificity of each of the three detailed investigation tasks by including more precise descriptions from MITRE ATT\&CK and some in-context examples about the target behavior; we did not find any common limitations or necessary clarifications for the classification task.
Third, we added additional instructions to elicit and produce multi-step reasoning and deliberation before outputting a final conclusion~\cite{NVIDIA_CoT_Prompting, lin2023just}.
Finally, we added a model persona (computer security expert) and modified the prompt to follow a markdown formatting~\cite{he2024does}. 
Appendix~\ref{sec:promptv2_template} shows the general template for the Version 2 prompts and our exfiltration task's prompt for this version.

\subsection{Evaluation Metrics}\label{sec:evaluation_metrics}
Using our groundtruth labels, our benchmark quantifies the performance of LLMs on our benchmark tasks by computing the \textbf{True Positive Rate (TPR)}, \textbf{False Positive Rate (FPR)}, and \textbf{F1 score}~\cite{wiki:Fscore}.
For each task, and a given data representation, the TPR equals the fraction of total attack instances correctly identified by the LLM across all scenarios in the task.
In most attack scenarios, there is only one attack instance in our groundtruth for each scenario; but some scenarios have multiple attack instances in a scenario (e.g., the Lateral Movement scenarios in our \optc derived data).
Our benchmark's automatic evaluator parses a model's output and compares the key identifiers for each task against the groundtruth (\S~\ref{sec:benchmark:tasks}) and counts how many unique groundtruth entities the model's output contains (divided by the total number of unique entities across a task).\footnote{Our evaluation framework uses both a standard JSON parser, as well as a custom parser designed to correctly handle and score LLM outputs that have errors in their JSON format. 
We manually verified that our framework's parsing correctly and automatically scores LLM responses across all scenarios and models we tested.}
Because our classification task consists of a binary label per scenario,
we score it asymmetrically: a model receives one true positive per attack scenario in which it flags \emph{any} chunk as malicious, while each chunk in a benign scenario that it flags as malicious counts as a separate false positive.

\begin{table*}[t]
\centering
\caption{Summary of LLM performances across our benchmark's \textbf{\lab} data (\S~\ref{sec:benchmark:data:lab}). 
Each row shows the performance metric value 
for the LLMs across a task and data representation (Column 2: ``Rep'').
We highlight and bold the best value for each row (highest F1 score, highest TPR, or lowest FPR). 
For the classification task, the false positive denominators reflect the number of log chunks; while for the other three tasks, the denominator reflects the total number of log lines (\S~\ref{sec:evaluation_metrics}).
}
\label{tab:llm_performance_summary_lab}
\setlength{\tabcolsep}{3pt}
\begin{tabular}{lcc|rrrrr}
\toprule
Task & Rep & Metric & \thead{Gpt-5-mini,\\ \prompttwo} & \thead{Gemini-2.5-flash,\\ \prompttwo} & \thead{Llama 4 Maverick,\\ \prompttwo} & \thead{Gpt-5,\\ \prompttwo} & \thead{Gemini-2.5-pro,\\ \prompttwo} \\
\midrule
classification & edge & F1 & 0.57 & 0.36 & \cellcolor{green!20}{\textbf{1.00}} & 0.77 & 0.36 \\
classification & edge & TPR & 2/5 (40.00\%) & \cellcolor{green!20}{\textbf{5/5 (100.00\%)}} & \cellcolor{green!20}{\textbf{5/5 (100.00\%)}} & \cellcolor{green!20}{\textbf{5/5 (100.00\%)}} & \cellcolor{green!20}{\textbf{5/5 (100.00\%)}} \\
classification & edge & FPR & \cellcolor{green!20}{\textbf{0/21 (0.00\%)}} & 18/21 (85.71\%) & \cellcolor{green!20}{\textbf{0/21 (0.00\%)}} & 3/21 (14.29\%) & 18/21 (85.71\%) \\
\hline
\lm & edge & F1 & 0.57 & 0.48 & \cellcolor{green!20}{\textbf{0.62}} & 0.33 & 0.45 \\
\lm & edge & TPR & 2/5 (40.00\%) & \cellcolor{green!20}{\textbf{5/5 (100.00\%)}} & 4/5 (80.00\%) & 1/5 (20.00\%) & \cellcolor{green!20}{\textbf{5/5 (100.00\%)}} \\
\lm & edge & FPR & \cellcolor{green!20}{\textbf{0/19950 (0.00\%)}} & 11/19950 (0.06\%) & 4/19950 (0.02\%) & \cellcolor{green!20}{\textbf{0/19950 (0.00\%)}} & 12/19950 (0.06\%) \\
\hline
persistence & edge & F1 & \cellcolor{green!20}{\textbf{0.80}} & 0.67 & 0.57 & 0.57 & 0.47 \\
persistence & edge & TPR & \cellcolor{green!20}{\textbf{4/5 (80.00\%)}} & \cellcolor{green!20}{\textbf{4/5 (80.00\%)}} & 2/5 (40.00\%) & 2/5 (40.00\%) & \cellcolor{green!20}{\textbf{4/5 (80.00\%)}} \\
persistence & edge & FPR & 1/8412 (0.01\%) & 3/8412 (0.04\%) & \cellcolor{green!20}{\textbf{0/8412 (0.00\%)}} & \cellcolor{green!20}{\textbf{0/8412 (0.00\%)}} & 8/8412 (0.10\%) \\
\hline
exfiltration & edge & F1 & 0.50 & 0.56 & 0.29 & \cellcolor{green!20}{\textbf{0.77}} & 0.40 \\
exfiltration & edge & TPR & 3/5 (60.00\%) & \cellcolor{green!20}{\textbf{5/5 (100.00\%)}} & 1/5 (20.00\%) & \cellcolor{green!20}{\textbf{5/5 (100.00\%)}} & \cellcolor{green!20}{\textbf{5/5 (100.00\%)}} \\
exfiltration & edge & FPR & 4/16448 (0.02\%) & 8/16448 (0.05\%) & \cellcolor{green!20}{\textbf{1/16448 (0.01\%)}} & 3/16448 (0.02\%) & 15/16448 (0.09\%) \\
\hline
classification & raw & F1 & \cellcolor{green!20}{\textbf{1.00}} & 0.75 & 0.80 & 0.80 & 0.46 \\
classification & raw & TPR & \cellcolor{green!20}{\textbf{3/3 (100.00\%)}} & \cellcolor{green!20}{\textbf{3/3 (100.00\%)}} & 2/3 (66.67\%) & 2/3 (66.67\%) & \cellcolor{green!20}{\textbf{3/3 (100.00\%)}} \\
classification & raw & FPR & \cellcolor{green!20}{\textbf{0/32 (0.00\%)}} & 2/32 (6.25\%) & \cellcolor{green!20}{\textbf{0/32 (0.00\%)}} & \cellcolor{green!20}{\textbf{0/32 (0.00\%)}} & 7/32 (21.88\%) \\
\hline
\lm & raw & F1 & 0.00 & 0.00 & 0.00 & 0.00 & \cellcolor{green!20}{\textbf{1.00}} \\
\lm & raw & TPR & 0/1 (0.00\%) & 0/1 (0.00\%) & 0/1 (0.00\%) & 0/1 (0.00\%) & \cellcolor{green!20}{\textbf{1/1 (100.00\%)}} \\
\lm & raw & FPR & 1/30425 (0.00\%) & 4/30425 (0.01\%) & \cellcolor{green!20}{\textbf{0/30425 (0.00\%)}} & \cellcolor{green!20}{\textbf{0/30425 (0.00\%)}} & \cellcolor{green!20}{\textbf{0/30425 (0.00\%)}} \\
\hline
persistence & raw & F1 & 0.57 & 0.60 & 0.33 & 0.50 & \cellcolor{green!20}{\textbf{1.00}} \\
persistence & raw & TPR & 2/5 (40.00\%) & 3/5 (60.00\%) & 1/5 (20.00\%) & 2/5 (40.00\%) & \cellcolor{green!20}{\textbf{5/5 (100.00\%)}} \\
persistence & raw & FPR & \cellcolor{green!20}{\textbf{0/34511 (0.00\%)}} & 2/34511 (0.01\%) & \cellcolor{green!20}{\textbf{0/34511 (0.00\%)}} & 1/34511 (0.00\%) & \cellcolor{green!20}{\textbf{0/34511 (0.00\%)}} \\
\hline
exfiltration & raw & F1 & 0.62 & 0.62 & 0.29 & \cellcolor{green!20}{\textbf{0.67}} & 0.57 \\
exfiltration & raw & TPR & \cellcolor{green!20}{\textbf{4/4 (100.00\%)}} & \cellcolor{green!20}{\textbf{4/4 (100.00\%)}} & 1/4 (25.00\%) & 3/4 (75.00\%) & \cellcolor{green!20}{\textbf{4/4 (100.00\%)}} \\
exfiltration & raw & FPR & 5/50267 (0.01\%) & 5/50267 (0.01\%) & \cellcolor{green!20}{\textbf{2/50267 (0.00\%)}} & \cellcolor{green!20}{\textbf{2/50267 (0.00\%)}} & 6/50267 (0.01\%) \\
\bottomrule
\end{tabular}
\end{table*}

\subsection{Limitations}\label{sec:limitations}

Our benchmark's dataset does not cover all possible attacks a security team might investigate, and it does not test attacks that aim to deliberately evade LLMs.
For our classification task, our benchmark also does not evaluate attacks where the necessary log information spans very long durations, due to the prohibitive cost for both researchers and practitioners to feasibly use LLMs for these cases.
Additionally, our benign scenarios may not capture the full space of all benign activity and false alarms in practice.
Nonetheless, our benchmark makes an important contribution by presenting a set of well-defined tasks for this problem and a novel dataset with detailed groundtruth.
Future work can continue to expand upon our dataset, which spans both newly generated data and carefully curated public data sources, multiple benign scenarios specifically designed to simulate false positives from a detector (\S~\ref{sec:benchmark:data:optc}), and dozens of different attack behaviors.

Our experiments (\S~\ref{sec:results}) do not test every possible LLM, prompt, or application architecture, since it is impossible to exhaustively cover the space.
Instead, we selected a reasonably representative set of the largest state-of-the-art models (at the time of our experiments) and smaller frontier models, and devised different prompts following prompt engineering best practices.
Additionally due to costs, our experiments evaluate the models using a temperature of 0 (when available) to achieve deterministic results.
Our experiments aim to provide a first look and new insights into the different subtleties and design considerations that developers and researchers should take into account when using/designing LLMs for security investigations.
We acknowledge that newer models have and will continue to be released. 
By open sourcing all our data and code (including model inputs/outputs), we provide a necessary foundation for future work to study this space and analyze the performance of new models.



\begin{table*}[t]
\centering
\caption{Summary of LLM performances on our benchmark's \textbf{\optc data} (\S~\ref{sec:benchmark:data:optc}). For the classification task, the false positive denominators reflect the number of log chunks; while for the other three tasks, the denominator reflects the total log lines (\S~\ref{sec:evaluation_metrics}).
} 
\label{tab:llm_performance_summary_optc}
\begin{tabular}{lcc|rrrrr}
\toprule
Task & Rep & Metric & \thead{Gpt-5-mini,\\ \prompttwo} & \thead{Gemini-2.5-flash,\\ \prompttwo} & \thead{Llama 4 Maverick,\\ \prompttwo} & \thead{Gpt-5,\\ \prompttwo} & \thead{Gemini-2.5-pro,\\ \prompttwo} \\
\midrule
classification & edge & F1 & \cellcolor{green!20}{0.45} & 0.08 & 0.29 & \cellcolor{green!20}{0.45} & 0.08 \\
classification & edge & TPR & \cellcolor{green!20}{5/5 (100.00\%)} & \cellcolor{green!20}{5/5 (100.00\%)} & \cellcolor{green!20}{5/5 (100.00\%)} & \cellcolor{green!20}{5/5 (100.00\%)} & \cellcolor{green!20}{5/5 (100.00\%)} \\
classification & edge & FPR & \cellcolor{green!20}{12/145 (8.28\%)} & 115/145 (79.31\%) & 24/145 (16.55\%) & \cellcolor{green!20}{12/145 (8.28\%)} & 116/145 (80.00\%) \\
\hline
\lm & edge & F1 & 0.18 & \cellcolor{green!20}{0.25} & 0.05 & 0.08 & 0.20 \\
\lm & edge & TPR & 3/19 (15.79\%) & \cellcolor{green!20}{7/19 (36.84\%)} & 1/19 (5.26\%) & 1/19 (5.26\%) & 6/19 (31.58\%) \\
\lm & edge & FPR & 12/61121 (0.02\%) & 30/61121 (0.05\%) & 15/61121 (0.02\%) & \cellcolor{green!20}{3/61121 (0.00\%)} & 33/61121 (0.05\%) \\
\hline
persistence & edge & F1 & 0.43 & 0.21 & 0.17 & \cellcolor{green!20}{0.60} & 0.20 \\
persistence & edge & TPR & 3/4 (75.00\%) & 3/4 (75.00\%) & 2/4 (50.00\%) & 3/4 (75.00\%) & \cellcolor{green!20}{4/4 (100.00\%)} \\
persistence & edge & FPR & 7/64234 (0.01\%) & 22/64234 (0.03\%) & 18/64234 (0.03\%) & \cellcolor{green!20}{3/64234 (0.00\%)} & 34/64234 (0.05\%) \\
\hline
exfiltration & edge & F1 & \cellcolor{green!20}{1.00} & 0.29 & 0.50 & \cellcolor{green!20}{1.00} & 0.15 \\
exfiltration & edge & TPR & \cellcolor{green!20}{2/2 (100.00\%)} & \cellcolor{green!20}{2/2 (100.00\%)} & \cellcolor{green!20}{2/2 (100.00\%)} & \cellcolor{green!20}{2/2 (100.00\%)} & \cellcolor{green!20}{2/2 (100.00\%)} \\
exfiltration & edge & FPR & \cellcolor{green!20}{0/59948 (0.00\%)} & 10/59948 (0.02\%) & 4/59948 (0.01\%) & \cellcolor{green!20}{0/59948 (0.00\%)} & 22/59948 (0.04\%) \\
\bottomrule
\end{tabular}
\end{table*}

\begin{figure*}[t]
\centering
\begin{subfigure}{0.98\columnwidth}
    \includegraphics[width=0.99\linewidth]{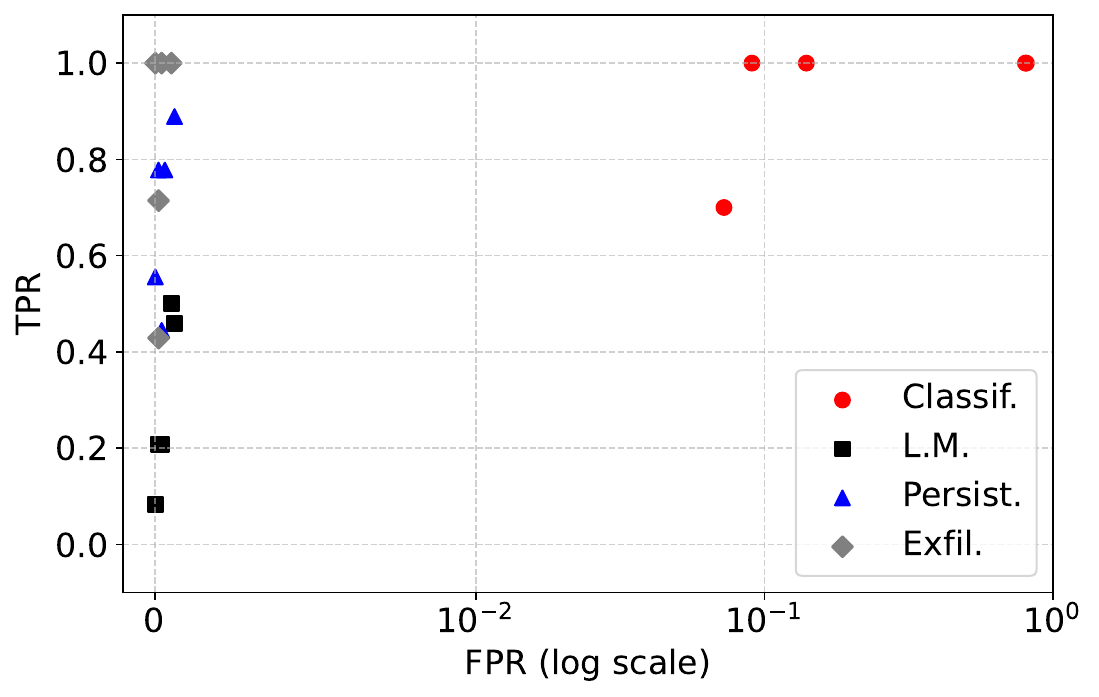}
    \caption{}
    \label{fig:task_comp_edge}
\end{subfigure}
\hfill
\begin{subfigure}{0.98\columnwidth}
    \includegraphics[width=0.99\linewidth]{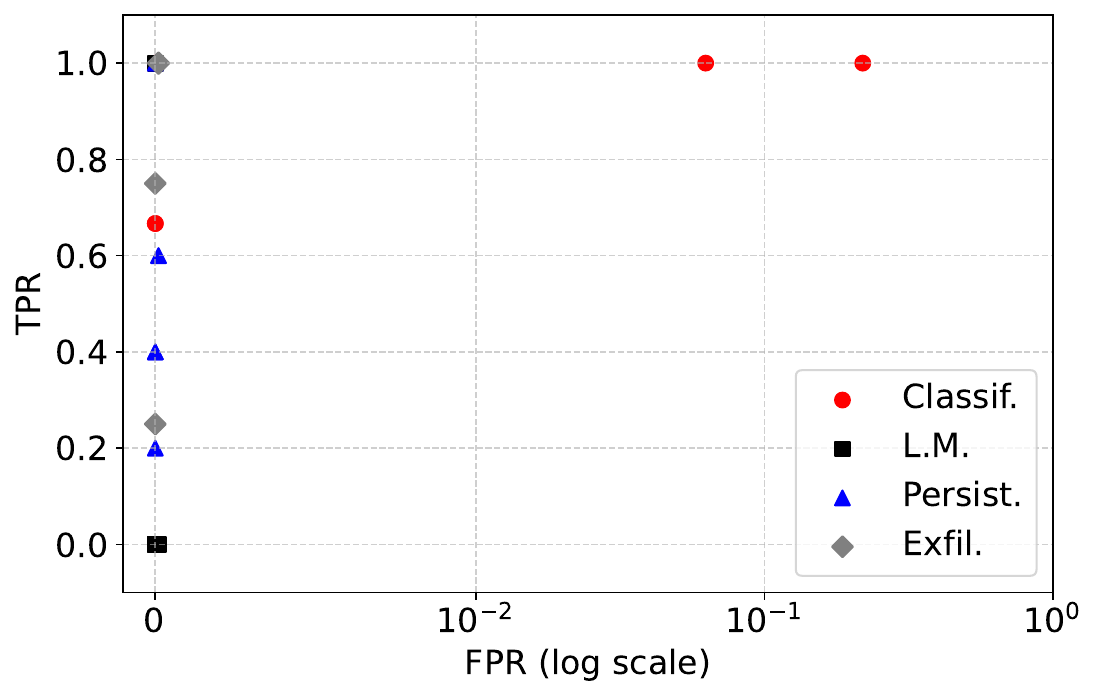}
    \caption{}
    \label{fig:task_comp_raw}
\end{subfigure}
\caption{Performance of the 5 different LLMs across the benchmark's tasks (\S~\ref{sec:results_overall}); each point represents one LLM on a task, where some points overlap due to equal performance. Fig (a) shows the performance on tasks given the \textbf{edge} representation, while Fig (b) shows the \textbf{raw} representation performance. The top left corner represents optimal performance, with lower false positive rates (x-axis) and higher true positive rates (y-axis). 
}
\label{fig:task_comparsion}
\end{figure*}

\section{Results and Analysis}\label{sec:results}


This section presents the results and analysis of evaluating five popular LLMs on our benchmark dataset.
First, we find that no model achieves strictly better performance across all tasks (\S~\ref{sec:results_overall}), with models frequently exhibiting overly-suspicious and false-positive prone verdicts across many tasks. 
Second, contrary to scaling expectations, smaller LLMs showed competitive and sometimes better performance than larger models (\S~\ref{sec:results_size}).
Third, using a more efficient edge data representation strictly improves performance for some models; however, we do see model-specific and task-specific effects (\S~\ref{sec:results_representation}).
Finally, changes to the prompts for a task can also have model-dependent effects (\S~\ref{sec:results_prompts}).
Thus, comparing performance between models requires careful exploration of different designs and configurations, rather than a one-size-fits-all approach.

\subsection{Benchmark Difficulty and Overall Performance Results}\label{sec:results_overall} 

Tables~\ref{tab:llm_performance_summary_lab} and~\ref{tab:llm_performance_summary_optc} show the performance of LLMs on the \lab and \optc portions of our dataset respectively. 
Overall, no model dominates across all tasks, where different models achieve the best F1, TPR, and FPR values depending on the task and data representation.

In terms of difficulty, our benchmark's data and tasks seem to present meaningful challenges for recent frontier models.
Only a few tasks have models with perfect performance (\eg classification on the \lab data and exfiltration on the \optc data); and even for these tasks, most models produce errors.
The \optc scenarios generally lead to worse performance on the classification, lateral movement, and persistence tasks, while the exfiltration scenarios in our \lab data seem more difficult compared to the task's \optc scenarios, with max F1-scores of 0.67 vs. 1.0 respectively.

Additionally, as seen in Figure~\ref{fig:task_comparsion}, 
model performance varies notably across tasks, although the relative difficulty of each task remains fairly consistent across LLMs.
LLMs performed the best on the exfiltration task, followed by the persistence task, where the top left of Figure~\ref{fig:task_comparsion}, with higher TPR and lower FPR, represent optimal performance.
In contrast, models tend to perform worst on average for the lateral movement task, which yields the lowest median F1-score across the five LLMs. 
Finally, most notably in the OpTC portion of our dataset, most models tend to produce fairly high false positive rates on both the lateral movement and classification tasks, with dozens of false positives across several thousands of log lines.


\subsection{Impact of Model Size}\label{sec:results_size}
Surprisingly, our experiments suggest that the larger LLMs do not always achieve the best results when compared to much smaller models.
At the time of our experiments, \gpt and \geminipro were two of the largest frontier LLMs, whereas \gptmini, \geminiflash, and \llamamav represent high-performing but smaller models. 
But as seen in Tables~\ref{tab:f1_score_small_large_labgen} and~\ref{tab:f1_score_small_large_optc}, on a specific task, the best-performing smaller model can achieve higher or equivalent performance than the best performing larger model when analyzing the edge representation of the data.
This competitiveness holds across both the \lab and \optc portions of our benchmark data.
For practitioners seeking to use LLMs to aid security investigations, this finding suggests that they might not need to use larger and more expensive models for optimal performance.

\begin{table}[t]
\begin{minipage}[t]{0.49\textwidth}
\centering
\captionsetup{width=0.99\columnwidth}
\caption{A simplified version of Tables~\ref{tab:llm_performance_summary_lab} and~\ref{tab:llm_performance_summary_optc} that compares the highest F1 score on our benchmark's \textbf{\lab data} across smaller models (\gptmini, \geminiflash, \llamamav) vs. larger models (\S~\ref{sec:results_size}). 
For the lateral movement task using raw logs, we have only one testable attack scenario, with one true positive (TP); so the max F1 score of 0 reflects all small models missing the TP.
}
\label{tab:f1_score_small_large_labgen}
\setlength{\tabcolsep}{4pt}
\begin{tabular}{lccrr}
\toprule
Task & Rep & Metric & \thead{Max F1 \\ (smaller models)} & \thead{Max F1\\ (larger models)} \\
\midrule
classification & edge & F1 & \cellcolor{green!20}{\textbf{1.00}} & 0.77 \\
\lm & edge & F1 & \cellcolor{green!20}{\textbf{0.62}} & 0.45 \\
persistence & edge & F1 & \cellcolor{green!20}{\textbf{0.80}} & 0.57 \\
exfiltration & edge & F1 & 0.56 & \cellcolor{green!20}{\textbf{0.77}} \\
classification & raw & F1 & \cellcolor{green!20}{\textbf{1.00}} & 0.80 \\
\lm & raw & F1 & 0.00 & \cellcolor{green!20}{\textbf{1.00}} \\
persistence & raw & F1 & 0.60 & \cellcolor{green!20}{\textbf{1.00}} \\
exfiltration & raw & F1 & 0.62 & \cellcolor{green!20}{\textbf{0.67}} \\
\bottomrule
\end{tabular}
\end{minipage}
\par
\begin{minipage}[t]{0.49\textwidth}
\centering
\captionsetup{width=0.99\columnwidth}
\caption{Comparisons of the best smaller model vs. larger model performance on our benchmark's \textbf{\optc data}. Overall smaller models often perform on-par with the largest frontier models (\S~\ref{sec:results_size}).}
\label{tab:f1_score_small_large_optc}
\setlength{\tabcolsep}{4pt}
\begin{tabular}{lccrr}
\toprule
Task & Rep & Metric & \thead{Max F1 \\ (smaller models)} & \thead{Max F1\\ (larger models)} \\
\midrule
classification & edge & F1 & \cellcolor{green!20}{\textbf{0.45}} & \cellcolor{green!20}{\textbf{0.45}} \\
\lm & edge & F1 & \cellcolor{green!20}{\textbf{0.25}} & 0.21 \\
persistence & edge & F1 & 0.43 & \cellcolor{green!20}{\textbf{0.60}} \\
exfiltration & edge & F1 & \cellcolor{green!20}{\textbf{1.00}} & \cellcolor{green!20}{\textbf{1.00}} \\
\bottomrule
\end{tabular}
\end{minipage}
\end{table}

\begin{table}[t]
\centering
\caption{
The effect of chunk sizes on Gemini 2.5 Flash and Gemini 2.5 Pro for the edge representation using \prompttwo (\S~\ref{par:chunk_size}). Across both models, our default choice of 400-line chunks performs commensurate or better than alternative sizes.
}
\label{tab:context-window-combined}
\resizebox{\columnwidth}{!}{ 
\begin{tabular}{lll|c|c|c|c}
\toprule
\textbf{Task} & \textbf{Model} & \textbf{Metric} & \textbf{200 lines} & \textbf{400 lines} & \textbf{800 lines} & \textbf{1600 lines} \\
\midrule
\multirow{2}{*}{Classification} & \multirow{2}{*}{\makecell{Gem 2.5\\Flash}} & TPR & 5/5 (100.00)    & 5/5 (100.00)    & 5/5 (100.00)    & 5/5 (100.00) \\
                                &                                     & FPR & 26/39 (66.67)   & 18/21 (85.71)   & 10/11 (90.91)   & 7/7 (100.00) \\
\midrule
\multirow{2}{*}{L.M.} & \multirow{2}{*}{\makecell{Gem 2.5\\Flash}} & TPR & 4/5 (80.00)     & 5/5 (100.00)    & 5/5 (100.00)    & 5/5 (100.00) \\
                                  &                                     & FPR & 25/19950 (0.13) & 11/19950 (0.06) & 8/19950 (0.04)  & 6/19950 (0.03) \\
\midrule
\multirow{2}{*}{Persistence} & \multirow{2}{*}{\makecell{Gem 2.5\\Flash}} & TPR & 5/5 (100.00)    & 4/5 (80.00)     & 3/5 (60.00)     & 3/5 (60.00) \\
                             &                                     & FPR & 3/8412 (0.04)   & 4/8412 (0.05)   & 5/8412 (0.06)   & 5/8412 (0.06) \\
\midrule
\multirow{2}{*}{Exfiltration} & \multirow{2}{*}{\makecell{Gem 2.5\\Flash}} & TPR & 2/5 (40.00)     & 5/5 (100.00)    & 3/5 (60.00)     & 3/5 (60.00) \\
                              &                                     & FPR & 10/16448 (0.06) & 8/16448 (0.05)  & 10/16448 (0.06) & 8/16448 (0.05) \\
\midrule
\multirow{2}{*}{Classification} & \multirow{2}{*}{\makecell{Gem 2.5\\Pro}} & TPR & 5/5 (100.00)    & 5/5 (100.00)    & 5/5 (100.00)    & 5/5 (100.00) \\
                                &                                   & FPR & 29/39 (74.36)   & 18/21 (85.71)   & 11/11 (100.00)  & 7/7 (100.00) \\
\midrule
\multirow{2}{*}{L.M.} & \multirow{2}{*}{\makecell{Gem 2.5\\Pro}} & TPR & 4/5 (80.00)     & 5/5 (100.00)    & 5/5 (100.00)    & 5/5 (100.00) \\
                                  &                                   & FPR & 17/19950 (0.09) & 12/19950 (0.06) & 6/19950 (0.03)  & 2/19950 (0.01) \\
\midrule
\multirow{2}{*}{Persistence} & \multirow{2}{*}{\makecell{Gem 2.5\\Pro}} & TPR & 5/5 (100.00)    & 4/5 (80.00)     & 5/5 (100.00)    & 5/5 (100.00) \\
                             &                                   & FPR & 10/8412 (0.12)  & 8/8412 (0.10)   & 8/8412 (0.10)   & 5/8412 (0.06) \\
\midrule
\multirow{2}{*}{Exfiltration} & \multirow{2}{*}{\makecell{Gem 2.5\\Pro}} & TPR & 2/5 (40.00)     & 5/5 (100.00)    & 2/5 (40.00)     & 2/5 (40.00) \\
                              &                                   & FPR & 19/16448 (0.12) & 15/16448 (0.09) & 14/16448 (0.09) & 4/16448 (0.02) \\
\bottomrule
\end{tabular}%
}
\end{table}

\subsection{Impact of Data Representations}\label{sec:results_representation} 

\begin{table*}[ht]
\centering
\caption{Comparisons of LLM performance when given edge vs. raw data representations (\S~\ref{sec:results_representation}). Green cells show the highest model performance per task (row) for each model. For this comparison, we only analyze scenarios which have both raw and edge representations.
}
\label{tab:rep_comparison_rawequivalent_f1scores}
\setlength{\tabcolsep}{4pt}
\begin{tabular}{lc|rr|rr|rr|rr|rr|c}
\toprule
 & \multicolumn{1}{c}{} & \multicolumn{2}{c}{\gptmini} & \multicolumn{2}{c}{\geminiflash} & \multicolumn{2}{c}{\llamamav} & \multicolumn{2}{c}{\gpt} & \multicolumn{2}{c}{\geminipro} & \multirow{2}{*}{\thead{Comparable\\ scenarios}} \\
Task & \multicolumn{1}{c}{Metric} & edge & \multicolumn{1}{r}{raw} & edge & \multicolumn{1}{r}{raw} & edge & \multicolumn{1}{r}{raw} & edge & \multicolumn{1}{r}{raw} & edge & \multicolumn{1}{r}{raw} &  \\
\midrule
classification & F1 & 0 & \cellcolor{green!20}{\textbf{1.00}} & 0.25 & \cellcolor{green!20}{\textbf{0.75}} & \cellcolor{green!20}{\textbf{1.00}} & 0.80 & 0.67 & \cellcolor{green!20}{\textbf{0.80}} & 0.25 & \cellcolor{green!20}{\textbf{0.46}} & 8 \\
\lm & F1 & \cellcolor{green!20}{\textbf{0}} & \cellcolor{green!20}{\textbf{0}} & \cellcolor{green!20}{\textbf{0.40}} & 0 & \cellcolor{green!20}{\textbf{1.00}} & 0 & \cellcolor{green!20}{\textbf{0}} & \cellcolor{green!20}{\textbf{0}} & \cellcolor{green!20}{\textbf{1.00}} & \cellcolor{green!20}{\textbf{1.00}} & 6 \\
persistence & F1 & \cellcolor{green!20}{\textbf{0.80}} & 0.57 & \cellcolor{green!20}{\textbf{0.67}} & 0.60 & \cellcolor{green!20}{\textbf{0.57}} & 0.33 & \cellcolor{green!20}{\textbf{0.57}} & 0.50 & 0.47 & \cellcolor{green!20}{\textbf{1.00}} & 10 \\
exfiltration & F1 & 0.55 & \cellcolor{green!20}{\textbf{0.62}} & 0.50 & \cellcolor{green!20}{\textbf{0.62}} & \cellcolor{green!20}{\textbf{0.33}} & 0.29 & \cellcolor{green!20}{\textbf{0.73}} & 0.67 & 0.35 & \cellcolor{green!20}{\textbf{0.57}} & 9 \\
\bottomrule
\end{tabular}
\end{table*}

Table~\ref{tab:rep_comparison_rawequivalent_f1scores} compares the performance of using different data representations, edge vs. raw (\S~\ref{sec:data_representation}), across the different models.
For each task, we only compare the scenarios that had feasible raw representations; \eg the \optc dataset does not provide raw logs and some of the \lab scenarios generated on Windows machines have prohibitively large volumes of raw log lines (\S~\ref{sec:benchmark}).

Overall, the best-performing data representation can depend on the specific model and task.
For example, \geminipro performs better when given the raw log data for all tasks.
However, for many other models, the more efficient edge representation leads to equal or better performance on many tasks; and for Llama 4, the edge representation strictly improves performance.
This novel finding suggests that future work and practitioners should explore how different data pre-processing and transformation approaches can improve not only costs, but also performance.


\paragraph{Chunk Size Sensitivity}\label{par:chunk_size}
Due to the max context windows of some models, our benchmark partitions audit logs into chunks of approximately 100,000 tokens for evaluation (\S~\ref{sec:evaluation_inputs}).
The Gemini 2.5 Flash and Pro models support context windows of up to 1M tokens.
To evaluate the impact of chunk size (log lines per chunk), we conducted a sensitivity analysis using both models across our benchmark. 
As summarized in Table~\ref{tab:context-window-combined} and Table~\ref{tab:context-window-raw-combined}, performance remains fairly stable across different chunk sizes.
Occasionally the smallest chunk size omits necessary context; but our default, middle chunk size avoids this problem (\eg 400 lines for edge representations).

\begin{table*}[ht]
\centering
\caption{Comparisons of the performance for each LLM between our two prompt versions (\S~\ref{sec:prompt_construction}). The edge representation rows in this table show the performance over all scenarios across the \lab and \optc datasets.
Highlighted green cells indicate the best (or equal) performance between the prompts for each model. We replaced undefined F1 scores (caused by a model detecting 0 true positives) with a value of 0; in particular, for raw log representations, the lateral movement task had only 1 attack scenario that could be feasibly tested.}
\label{tab:prompt_version_comparison}
\setlength{\tabcolsep}{4pt}
\resizebox{\textwidth}{!}{%
\begin{tabular}{lcc|rr|rr|rr|rr}
\toprule
Task & Rep & Metric & \thead{Gpt-5-mini,\\ Prompt v1} & \thead{Gpt-5-mini,\\ Prompt v2} & \thead{Gemini-2.5-flash,\\ Prompt v1} & \thead{Gemini-2.5-flash,\\ Prompt v2} & \thead{Llama 4 Maverick,\\ Prompt v1} & \thead{Llama 4 Maverick,\\ Prompt v2} & \thead{Gemini-2.5-pro,\\ Prompt v1} & \thead{Gemini-2.5-pro,\\ Prompt v2} \\
\midrule
classification & edge & F1 & \cellcolor{green!20}{\textbf{0.48}} & \cellcolor{green!20}{\textbf{0.48}} & \cellcolor{green!20}{\textbf{0.22}} & 0.13 & 0.17 & \cellcolor{green!20}{\textbf{0.45}} & \cellcolor{green!20}{\textbf{0.14}} & 0.13 \\
\lm & edge & F1 & 0.20 & \cellcolor{green!20}{\textbf{0.24}} & 0.21 & \cellcolor{green!20}{\textbf{0.31}} & 0.06 & \cellcolor{green!20}{\textbf{0.21}} & \cellcolor{green!20}{\textbf{0.29}} & 0.28 \\
persistence & edge & F1 & 0.12 & \cellcolor{green!20}{\textbf{0.58}} & 0.11 & \cellcolor{green!20}{\textbf{0.34}} & 0.08 & \cellcolor{green!20}{\textbf{0.26}} & 0.09 & \cellcolor{green!20}{\textbf{0.27}} \\
exfiltration & edge & F1 & 0.12 & \cellcolor{green!20}{\textbf{0.62}} & 0.13 & \cellcolor{green!20}{\textbf{0.44}} & 0.04 & \cellcolor{green!20}{\textbf{0.40}} & 0.14 & \cellcolor{green!20}{\textbf{0.27}} \\
classification & raw & F1 & \cellcolor{green!20}{\textbf{1.00}} & \cellcolor{green!20}{\textbf{1.00}} & \cellcolor{green!20}{\textbf{1.00}} & 0.75 & \cellcolor{green!20}{\textbf{1.00}} & 0.80 & \cellcolor{green!20}{\textbf{0.55}} & 0.46 \\
\lm & raw & F1 & \cellcolor{green!20}{\textbf{0.67}} & 0.00 & \cellcolor{green!20}{\textbf{0.17}} & 0.00 & \cellcolor{green!20}{\textbf{0.00}} & \cellcolor{green!20}{\textbf{0.00}} & \cellcolor{green!20}{\textbf{1.00}} & \cellcolor{green!20}{\textbf{1.00}} \\
persistence & raw & F1 & 0.56 & \cellcolor{green!20}{\textbf{0.57}} & 0.36 & \cellcolor{green!20}{\textbf{0.60}} & \cellcolor{green!20}{\textbf{0.44}} & 0.33 & 0.24 & \cellcolor{green!20}{\textbf{1.00}} \\
exfiltration & raw & F1 & 0.29 & \cellcolor{green!20}{\textbf{0.62}} & 0.35 & \cellcolor{green!20}{\textbf{0.62}} & 0.06 & \cellcolor{green!20}{\textbf{0.29}} & 0.28 & \cellcolor{green!20}{\textbf{0.57}} \\
\bottomrule
\end{tabular}
}
\end{table*}

\subsection{Effects of Prompt Variations}\label{sec:results_prompts} 

So far, we used the \prompttwo variation of our prompts (\S~\ref{sec:prompt_construction}) when presenting results.
In this prompt version, the final verdict for each event can be one of three labels: ``low'', ``medium'', or ``high'' suspicion.
For our experimental results, we only considered entities with ``high'' suspicion verdicts as entities the model flags as an attack.
Events where the model assigns a ``low'' or ``medium'' verdict are treated as events the model considers benign (either true negatives or false negatives), along with anything the model doesn't report.
In this section, we explore the impact of these two choices: the change in performance if we use Version 1 of our prompt and the impact of adopting a more aggressive verdict interpretation (\eg treating ``medium suspicious'' events as activity flagged by the LLM).


\paragraph{Prompt Versions} Table~\ref{tab:prompt_version_comparison} compares the performance of using \promptone versus \prompttwo across all scenarios in our benchmark for four LLMs. (In these experiments, we omit \gpt due to costs, but our analysis still includes \gptmini and a different large LLM: \geminipro).
In most cases, \prompttwo leads to better results. 
For the persistence and exfiltration tasks, this second prompt version led to strictly better F1-scores for almost all models and both data representations. 
Additionally, for two smaller LLMs (\gptmini and \llamamav), the second prompt leads to optimal performance on all tasks when given the data's edge representation.
In all these cases, the F1-score increases due to reductions in the false positive rates, with often no change to the true positive rates.

We attribute most of \prompttwo's gains to its added task-specific specificity (e.g. specifying known attacker behavior like cloud uploads, HTTPS exfiltration, pre-exfiltration compression for exfiltration; and emphasizing suspicious \emph{outgoing} connections rather than DNS/CDN traffic for lateral movement) and explicitly scoping out common false-positive sources (e.g., system-generated and temporary files that we considered not-sensitive in our specific investigation setup for exfiltration). 
By providing these specific details, we aim to incorporate domain knowledge into the model's computation, which might not be well-defined or internalized during an LLM's existing training or RLHF instruction tuning.

However, we note that prompt changes do not always have uniform effects across all models. 
For example, on the classification task when using the edge representation, \prompttwo improves Llama 4's performance significantly, but leads to worse performance on \geminiflash. 
This mixed effect suggests that prompts can have noticeable model-dependent impacts, and that a one-size-fits-all-approach may not accurately compare models head-to-head.
\paragraph{Aggressive Verdict Interpretations}\label{sec:prompt2_defensive}
Our results so far use a conservative interpretation of the model's verdict (treating only ``high suspicious'' as attack outputs). However, analysts could go with a more aggressive interpretation that also treats ``medium'' suspicion as attacks, achieving higher true positives at the cost of more false positives.
We evaluated the impact of switching to this more aggressive prompt, and found this change leads to strictly worse performance for most models across almost all tasks and data representations; true positive rates often remained the same, while the volume of false positives increased. 
However, we did observe some rare model-specific effects with this design choice; 
unlike other models, \gpt had almost uniformly better performance when analyzing the raw audit log data with this more aggressive prompt version.
Table~\ref{tab:v3prompt_comparison_with_gpt5_lab:full} in the Appendix provides more granular details.

\
\section{Analysis}\label{sec:analysis}

In this section, we analyze the explanations in the models' outputs to better understand both the quality of LLM-generated explanations for true positives (Section~\ref{sec:explain-quality}) and the thematic characteristics in the errors that the LLMs made on investigation tasks (Section~\ref{sec:errors}).


\subsection{Analysis of LLM-generated explanations}\label{sec:explain-quality}
As described in Section~\ref{sec:prompt_construction}, our prompt instructs each LLM 
to generate a \verb|"deliberation"| field before producing its final 
verdict.
If accurate, these explanations can help analysts interpret why a model flagged an activity as highly suspicious, and facilitate faster and more accurate investigations.
In this subsection, we evaluate the correctness of LLM-generated explanations across true-positive activity they identify across all our models and tasks.

\textbf{Measuring Explanation Quality:}
To compute explanation quality, we treat the problem as a Natural Language Inference (NLI) task~\cite{dagan2005pascal}, also known as textual entailment, and apply an LLM-as-a-judge for assessing textual entailment. 
Here the  \textit{hypothesis} is the LLM-generated explanation, and the \textit{premise} is the ground truth attack description.
An explanation (hypothesis) is entailed if the text of the ground truth (premise) semantically supports its claims (about existence of specific attack behaviors and artifacts).
To construct each scenario's entailment ``premise'', our benchmark uses Gemini 2.5 Pro to produce a single paragraph that combined the ATT\&CK technique description with the detailed description of the list of attack steps we performed and recorded during our scenario generation (which includes details like timestamps, execution commands, etc.).
To facilitate fine-grained explanation assessment, we compute entailment scores for any chunk with true positives by treating the chunk's LLM-generated explanation as the entailment hypothesis.
Our benchmark computes quality (entailment) scores on a five-point Likert scale~\cite{zheng2023judging}, where 1 indicates no entailment (poor quality) and 5 indicates complete entailment (high quality).


We then used Gemini 2.5 Flash as a LLM judge with a cybersecurity expert persona to score each premise-hypothesis pair (the quality of each chunk's explanation across true positives for each model and all scenarios). 
The judge prompt instructs the LLM to increase the score when the LLM explanations contain legitimate attack relevant behaviors and low-level artifacts (\eg malicious file/process names).
In contrast, the prompt instructs the LLM-judge to penalize the score when the explanation contradicts the attack or introduces irrelevant details about benign activity (\eg PIDs or file names that do not relate to any attack activity).

\textbf{Explanation Quality Results: }
As seen in Table~\ref{tab:judge_evaluation}, LLMs have fairly good explanation quality across the benchmark for true positives they correctly identify.
Explanations for the classification task have a lower average entailment score ($3.79/5$) than those for lateral movement ($4.64/5$), persistence ($4.42/5$), and exfiltration ($4.23/5$). 


To validate the accuracy of these LLM-judged quality scores,
two authors manually rated a random sample of explanations (30 premise-hypothesis pairs) using the same 1-5 scale. 
Human raters achieved a weighted Cohen's $\kappa$ of 0.409, while the judge reached $\kappa$ of 0.575 against the human consensus; both comparisons had 93.3\% adjacent agreement (within $\pm1$ point). 
On average, the LLM-judge approach in our framework errs towards giving lower quality ratings (-0.2 points) than the human raters.
Overall, these numbers suggest our framework's automated approach to computing explanation quality is reasonably reliable.

\begin{table}[t]
\centering
\captionsetup{width=0.99\columnwidth}
\caption{Entailment scores of model-generated explanations across the benchmark tasks, on a scale of 1~--~5, where a score of 5 represents the highest explanation accuracy relative to the groundtruth (\S~\ref{sec:explain-quality}).}
\resizebox{\columnwidth}{!}{%
\begin{tabular}{lcccc}
\toprule
 & \textbf{Classification} & \textbf{Lat. Movement} & \textbf{Persistence} & \textbf{Exfiltration} \\
\midrule
\textbf{gemini-2.5-flash} & 3.34 & 4.80 & 4.12 & 3.96 \\
\textbf{gemini-2.5-pro}   & 3.36 & 4.40 & 4.00 & 4.00 \\
\textbf{gpt-5}            & 4.00 & 5.00 & 4.50 & 4.52 \\
\textbf{gpt-5-mini}       & 4.00 & 4.00 & 4.50 & 3.67 \\
\textbf{llama-4-maverick} & 4.27 & 5.00 & 5.00 & 5.00 \\
\midrule
\textbf{Avg. LLM score}          & 3.79 & 4.64 & 4.42 & 4.23 \\
\bottomrule
\end{tabular}%
}
\label{tab:judge_evaluation}
\end{table}

\subsection{Model Error Analysis}\label{sec:errors}
In this section, we examine the extent to which errors overlap across models and the characteristics of common errors.
We focus specifically on false positive errors, since they constitute the larger class of errors (Tables~\ref{tab:llm_performance_summary_lab} and~\ref{tab:llm_performance_summary_optc}), and represent one of the biggest challenges and bottlenecks for security teams in practice~\cite{alahmadi202299}.
Additionally, false positive errors produce concrete entities that we can compare across models, and include specific reasoning.
In contrast, false negatives by their nature often have an absence of output or specific suspicious entities (since models consider all activity to be benign), which makes it difficult to study false negative errors.


\begin{figure}[t]
\centering
\includegraphics[width=0.8\columnwidth]{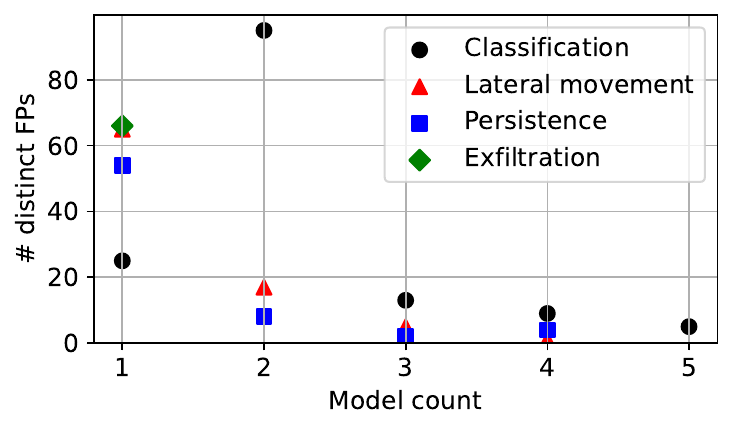}
\caption{The number of distinct false positives (y-axis) that exactly N models generate (x-axis), for each task across our benchmark when using the data's edge representation. These numbers include all false positives, across both the \lab and \optc portions of our benchmark.}
\label{fig:error_scatter}
\end{figure}

Figure~\ref{fig:error_scatter} shows the number of distinct false positives (FPs) generated by exactly 1, 2, 3, 4, or all 5 models, across all scenarios for each task in our benchmark when using the edge representation of audit logs.
For the most part, the same error rarely gets flagged by a majority of models: a total of 350 unique FPs across all tasks only occur for 1 or 2 models.
When processing the edge representation, only 19 FPs get produced by four or more models (14 false positives from the classification task, 1 from the lateral movement task, and 4 from the persistence task).
For scenarios evaluated on the raw log representation (not shown in the figure), we observed even smaller error overlap (due in part to the smaller set of scenarios with available raw logs).
Among the 37 unique false positives across all tasks, none were flagged by three or more models, and only 1 false positive was produced by two models.
% ...

Additionally, we saw minimal overlap between the errors produced by models when given two different data representations of the same scenario.
Among 75 total FPs from all models on all scenarios with both raw and edge representations, only one FP (in a persistence scenario) was common across the errors generated from processing the edge representation and when processing the raw logs.
This lack of overlap suggests distinct differences in how LLMs interpret security logs, and provides potential opportunities to defenders.
For example, practitioners may be able to use ensembles of different models to achieve lower false positive rates.

\subsection{Characteristics of Common Errors}\label{sec:error_char}
To understand the reasons for common errors, we manually investigated the 19 ``common'' false positives generated by 4 or more LLMs; all of these errors come from the \optc data in our benchmark. 
Our prompts instruct the models to output reasoning traces for their decisions (\eg in output fields that ask an LLM to specify the ``key log lines'', ``evidence'', and a ``deliberation'' before the model outputs its verdicts).
We carefully examined these reasoning-related fields (and verified their factual consistency with the input logs) to better understand the characteristics of common errors.


When analyzing our LLMs' reasoning outputs for common false positives, we observed that the text consistently relied on two characteristics about the logging data:
\begin{itemize}
    \item The names of system objects in the logs, such as the names of processes, scripts, and/or tasks executed.
    \item The sequence of events in the logs. In some cases, the reasoning traces also highlight the volume of an activity in a short time window as a suspicious sign.
\end{itemize}

Concretely, 18 / 19 of the common false positives have a heavy reliance on system object names in the reasoning traces of many models.
In one common classification task false positive, \gptmini's deliberation highlights ``non-standard" names in the logs, such as \path|C:\\ncr\\outlook\\invis.vbs| (name of a script), \path|login_keys_64bit.bat|, \path|popup_keys_64bit.bat| (names of batch files that were executed), as suspicious signs.
Additionally, the model states that these ``names strongly suggesting credential or key-capture behavior".
As another example, for a common false positive among the persistence scenarios, \geminipro's deliberation states: ``The most damning evidence is the execution of a scheduled task explicitly named \path|\\clear_logs|". For this same FP instance, \gpt stated the following in its deliberation: ``a scheduled task named \path|\\\\clear_logs| runs a SYSTEM-level batch file at \path|C:\\ncr\\DeleteArchiveSecurity.bat|, whose name implies deletion of security log archives". This output clearly shows \gpt's reliance on the name of the batch file.

Similarly, across 17 of the 19 common false positives, the sequence of events appears frequently in model's explanation-related outputs.
In these cases, the deliberation field frequently mentioned terms such as ``sequence'', ``chain'', ``execution flow'', and referenced parent nodes and temporal relationships between entities. 
For instance, the deliberation sentence of \geminiflash on one of the common classification task false positives states:
``The fact that this script is executed by `cmd.exe' processes spawned by `svchost.exe', `csrss.exe', and `schtasks.exe' with SYSTEM privileges is highly alarming."
In another false positive instance, \geminiflash reports that ``The core of the suspicious activity lies in the \textit{execution chain}: `Explorer.EXE' $\to$ `WScript.exe' $\to$ `invis.vbs' and `login\_keys\_64bit.bat'".
It then states that this sequence is a key characteristic for attacks such as ``credential/keylogging or stealthy post-compromise scripting".

When manually analyzing these common false positives, we identified four false positives corresponding to security relevant activity that could be difficult to unambiguously label without external context. The activities included disabling system integrity checks, modifying Outlook anti-spam settings, and a powershell-launched outbound TCP connection. But they all occur in scenarios created from the DARPA \optc dataset, and from portions of that dataset explicitly labeled as benign by the official \optc distributors.
We respect the original DARPA labels but we recognize that these cases might represent situations where different security teams may want to interpret and act on the activity differently. Such ambiguity will inevitably occur in real-world logs across organizations. 
The publicly released data and code for our benchmark include the groundtruth labels, documentation describing our labeling process, and the functions for computing evaluation metrics. As a result, users and future work can modify this decision depending on their situation and goals.
Additionally, we carefully constructed the 25 \lab scenarios in our data to avoid any potentially ambiguous activity.

\section{Discussion and Conclusion}\label{sec:conclusion}

In this work, we developed a novel benchmark for evaluating the use of LLMs in security investigations. 
Our benchmark covers four common security investigation tasks, consisting of newly generated data and curated data from the most complex, publicly available security audit log dataset. 
We believe our benchmark serves as a useful measurement tool for researchers and practitioners to evaluate and improve LLMs' capabilities at security investigations.
As an example, we distill several interesting findings based on using our benchmark to analyze the performance of five popular LLMs.

First, our evaluation suggests that models appear to have an overly-suspicious interpretation of audit log data, incurring high numbers of false positives for some tasks (\S~\ref{sec:results_overall}).
Although some models can achieve both high true positives and low false positives when identifying exfiltration behavior, the LLMs we test frequently report many benign activities as potential attacks across other tasks.
Based on the explanation outputs from LLMs, these false positives often appear to over-index on potentially unusual names in files, processes, and other objects, as well as high-volume activities (\S~\ref{sec:error_char}).
Future work should explore ways to improve the robustness of models' understanding of attack behavior and reduce false positives. 
Given the small overlap in errors, one potential direction for mitigating false positives may be to explore using an ensemble of models to produce alerts (\S~\ref{sec:errors}).

Second, our results point to several directions for improving both the cost and performance of using LLMs for security investigations.
For example, we found that for many tasks, smaller models can perform just as well as larger models (\S~\ref{sec:results_size}).
Additionally, our work finds that for certain models, including the smaller, open-weight Llama 4 model, using a transformed and more efficient representation of log data leads to strictly better performance (\S~\ref{sec:results_representation}).
These findings indicate that data representation engineering, and combining insights from existing work on provenance graphs and information flow with LLMs could be promising avenues for future work, as ways to improve the cost, efficiency, and performance of LLMs on security investigation tasks.




Third, our experiments provide cautionary advice for comparing or choosing between specific LLMs to use.
In particular, our evaluations show that various design decisions can lead to model-specific performance changes.
For example, while prompt changes that improve the performance of one model tend to improve the performance of all models (\S~\ref{sec:results_prompts}), we observed some deviations from this trend across our benchmark.
Similarly, the transformed edge representation of our data led to strict improvements for some models, but worse performance for others.
Thus, we believe that practitioners and researchers need to carefully explore a range of different configurations and designs, and weigh the optimal choice for each model when making cross-model comparisons in practice.
Our datasets and benchmark provide the means to perform this assessment and decision making.

Ultimately, this paper contributes both an extensible benchmark and novel insights for security teams, model developers, and researchers exploring how to use and improve LLMs for security investigations. 
We publicly release all our code, data, documentation, prompts, and model input/outputs as part of our benchmark contribution.
We hope that future work continues to use and extend our benchmark as a means to study and improve the use of LLMs for investigating security audit logs.

\bibliographystyle{ACM-Reference-Format}
\bibliography{references}

\appendix

\section{Dataset Details}\label{sec:appendix:data}

\subsection{Description of \lab scenarios} \label{sec_appendix_dataset_lab}

\paragraph{Classification (Attack Scenarios)}
\begin{itemize}
    \item Scenario 1: 
    T1563.002 Remote Service Session Hijacking, followed by validation using visudo and appending insecure default which disables the requirement for user must re-auth.
    \item Scenario 2: 
    TA0005 Defense evasion - T1548.003 Abuse Elevation Control Mechanism: Sudo and Sudo Caching
    
    \item Scenario 3: 
    T1555.003 Credentials from Password Stores: Credentials from Web Browsers and save the extracted credentials in a file for exfiltration.
    \item Scenario 4: 
    TA0040 Impact - T1486 Data Encrypted for Impact, followed by deletion of original file and keys
    \item Scenario 5: 
    A multi-stage attack chain. 
    MITRE Techniques: TA0001 Initial access - T1566.002 Phishing: Spearphishing Link, TA0005 Defense evasion - 1562.001 Impair Defenses: Disable or Modify Tools, TA0002 Execution - T1059.005 Command and Scripting Interpreter: Visual Basic, TA0010 Exfiltration - T1041 Exfiltration Over C2 Channel
\end{itemize}

\paragraph{Benign Scenarios}
\begin{itemize}
\item Benign Scenario 1: A user accesses a remote Ubuntu VM's GUI via VNC (requiring an SSH tunnel) to perform email management (sending/reading emails).
\item Benign Scenario 2: Programming activity. A user writes and executes Python scripts using the terminal.
\item Benign Scenario 3: A user accesses a cloud-hosted Ubuntu VM's GUI, opens applications (e.g., File Manager, web browser), closes the GUI session once, reopens it, and continues web browsing.
\item Benign Scenario 4: Mixed desktop and web activities. A user navigates the file system (terminal), browses the web (Brave), manages emails (Thunderbird), and plays Sudoku.
\item Benign Scenario 5: System administration and file management. A user performs routine system maintenance (package updates, storage monitoring, file creation, process inspection) using command-line tools.
\end{itemize}

\paragraph{Lateral Movement}
\begin{itemize}
    \item Scenario 1: 
    T1563.002 Remote Service Session Hijacking
    \item Scenario 2: Credential reuse attack.
    T1021.001 Remote Services: Remote Desktop Protocol
    \item Scenario 3: Secure shell lateral access.
    T1021.004 Remote Services: SSH
    \item Scenario 4: Centralized management exploitation.
    T1072 Software Deployment Tools
    \item Scenario 5: Hash-based authentication bypass.
    T1550.002 Use Alternate Authentication Material: Pass the Hash
\end{itemize}

\paragraph{Persistence}
\begin{itemize}
    \item Scenario 1: SSH authentication bypass.
    Technique: T1098.004 Account Manipulation: SSH Authorized Keys
    \item Scenario 2: Local account creation.
    Technique: T1136.001 Create Account: Local Account
    \item Scenario 3: Reverse SSH tunnel.
    Technique: T1133 External Remote Service
    \item Scenario 4: System startup modification.
    Technique: T1037.004 Boot or Logon Initialization Scripts: RC Scripts
    \item Scenario 5: Automated task scheduling.
    Technique: T1053.003 Scheduled Task/Job: Cron
\end{itemize}

\paragraph{Data Exfiltration}
\begin{itemize}
    \item Scenario 1: HTTPS protocol data transfer.
    T1048.002 Exfiltration Over Alternative Protocol: Exfiltration Over Asymmetric Encrypted Non-C2 Protocol
    \item Scenario 2: Command and control channel exfiltration.
    T1041 Exfiltration Over C2 Channel
    \item Scenario 3: Cloud service data extraction.
    T1567.002 Exfiltration Over Web Service: Exfiltration to Cloud Storage
    \item Scenario 4: DNS protocol data transfer.
    T1048.003 Exfiltration Over Alternative Protocol: Exfiltration Over Unencrypted/Obfuscated Non-C2 Protocol
    \item Scenario 5: Physical media data transfer.
    T1052.001 Exfiltration Over Physical Medium: Exfiltration over USB
\end{itemize}

\subsection{Description of \optc scenarios} \label{sec:appendix:dataset:optc}

\paragraph{Classification (Attack Scenarios)}
\begin{itemize}
    \item 
    Scenario 1: Memory credential extraction.
    Attacker extracts plaintext passwords from system memory and bypasses UAC via Group Policy manipulation.
    \item 
    Scenario 2: Network reconnaissance via Ping Sweep.
    Attacker runs automated scripts to discover active hosts across network subnet ranges.
    \item 
    Scenario 3: UAC bypass and credential search.
    Attacker exploits Windows Event Viewer for privilege escalation and searches Group Policy Objects for stored credentials.
    \item 
    Scenario 4: System profiling and process migration.
    Attacker uses remote access tool modules to discover system/network info (apps, domain, shares) and migrates to critical system processes.
    \item 
    Scenario 5: Advanced memory credential extraction.
    Attacker deploys specialized credential extraction tools while maintaining external command and control (C2) communication. 
\end{itemize}

\paragraph{Lateral Movement}
\begin{itemize}
    \item Scenario 1: Lateral access based on WMI.
    Attacker uses Windows Management Instrumentation (WMI) to remotely execute commands and establish access on a target system.
    \item Scenario 2: Lateral access based on administrative WMI.
    Attacker leverages administrative credentials via WMI to pivot from a compromised system to another network host.
    \item Scenario 3: Lateral access based on Powershell.
    Attacker injects a PowerShell script to establish an encrypted communication channel with an external C2.
    \item Scenario 4: Lateral access based on WMI with domain controller targeting.
    Attacker uses WMI to pivot from a compromised machine to a high-value domain controller.
    \item Scenario 5: Lateral access based on an RPC-based network.
    Attacker performs systematic lateral movement across network segments using Remote Procedure Call connections on administrative ports.
\end{itemize}

\paragraph{Persistence}
\begin{itemize}
    \item Scenario 1: 
    T1547.001 Boot or Logon Autostart Execution: Registry Run Keys / Startup Folder.
    \item Scenario 2: 
    T1546.003 Event Triggered Execution: Windows Management Instrumentation Event Subscription, T1053.005 Scheduled Task/Job: Scheduled Task.
    \item Scenario 3: 
    T1547.001 Boot or Logon Autostart Execution: Registry Run Keys / Startup Folder.
    \item Scenario 4: 
    T1133 External Remote Services.
\end{itemize}

\paragraph{Data Exfiltration.}
\begin{itemize}
    \item Scenario 1: Command and control channel data exfiltration.
    Attacker compresses the documents directory into a ZIP archive and exfiltrates it to an external C\&C server using a network transfer utility.
    \item Scenario 2: Alternative protocol data exfiltration.
    Attacker compresses network share files into a multi-gigabyte archive and exfiltrates it to an external server using a custom transfer tool over an RDP session.
\end{itemize}

\section{Prompts for Investigation Tasks} \label{sec:appendix:prompt}
\subsection{\promptone template}
\begin{tcolorbox}[
    breakable,
    width=\linewidth,
    colback=gray!10,
    colframe=gray!50,
    boxrule=0.1pt,
    arc=1mm,
    boxsep=1mm,
    left=1mm,
    right=1mm,
    top=1mm,
    bottom=1mm,
    before skip=0pt,
    after skip=1ex
]
\small
You will be given a set of [Linux / Windows] system logs and need to [SECURITY TASK DEFINITION]. 
Examine ALL the lines of the audit log input.

\par
Task: [\textit{task description}]

\par
Output Format: [\textit{output format description}]

\par
Input data description \textbf{(for Linux raw logs)}:
The input audit log data of Linux system obtained directly from Auditd framework that should be used for [task] is as follows: {\textit{input data}}
\par
OR
\par
Input data description \textbf{(for Linux edge representation)}:
The audit log data from a Linux system has been processed into an information-flow graph.
Each line of input now represents an information flow edge that describes an action between two nodes, and each node corresponds to one of three objects: a process, file and network socket.
Each input edge contains the following information:
\texttt{'(timestamp of action) [unique source node ID] \{\{key-value pairs of source node attributes separated by commas\}\} $\rightarrow$ [unique edge ID] \{\{key-value pairs of the action's attributes separated by commas\}\} $\rightarrow$ [unique destination node ID] \{\{key-value pairs of destination node separated by commas\}\}'}.

The input audit log data that should be used for [task] is as follows: \\{[\textit{input data}]}
\end{tcolorbox}

\subsection{\promptone for Exfiltration Task}

\begin{tcolorbox}[
    breakable,
    width=\linewidth,
    colback=gray!10,
    colframe=gray!50,
    boxrule=0.1pt,
    arc=1mm,
    boxsep=1mm,
    left=1mm,
    right=1mm,
    top=1mm,
    bottom=1mm,
    before skip=1ex,
    after skip=1ex
]
\small
You will be given a set of Windows system logs and need to identify any data exfiltration actions where the user's data is stolen from the machine.
Examine ALL the lines of the audit log input.

\par
Task: For each relevant action in the logs that indicates data exfiltration behavior, return the timestamp, the PID of the process that performs exfiltration, full file path of the files that are stolen (to generate the full file path, combine the directory path and file name), the key log lines showing the data exfiltration of the files, and an explanation of the exfiltration activity.
Exfiltration actions consist of leaking, copying, or transferring files to external sources, causing data to leave the original system.
If you cannot identify any data exfiltration with clear evidence, do not make anything up.

\par
Output format:
Output the data exfiltration instances that you identify as a list of JSON, with each JSON object representing an exfiltration instance following this format:

\par
\#\#\#\#

\{``timestamp'': \{string, timestamp of the exfiltration action\},

``pid'': \{int, the PID of the process doing the data exfiltration action\},

``exfiltrated data'': \{string, the full file path of the files that are exfiltrated\},

``key log lines'': \{string, list of comma separated log lines showing an exfiltration action and all the data that is exfiltrated by the process as part of this exfiltration action, where every log line should be on a new line\},

``explanation'': \{string, describe all the evidence from the key log lines showing the files that are being exfiltrated. Based on the evidence, write a claim about whether or not this should be a data exfiltration instance\}
\}

\par
\#\#\#\#

\par
If no data exfiltration activity is found, return an empty JSON object.

\par
[Input data template for raw / edge representation]
\end{tcolorbox}

\subsection{\prompttwo template}\label{sec:promptv2_template}
\begin{tcolorbox}[
    breakable,
    width=\linewidth,
    colback=gray!10,
    colframe=gray!50,
    boxrule=0.1pt,
    arc=1mm,
    boxsep=1mm,
    left=1mm,
    right=1mm,
    top=1mm,
    bottom=1mm,
    before skip=1ex,
    after skip=1ex
]
\small
\textit{\{self.PERSONA\}}

\par
\textbf{Goal}

\par
\textit{\{self.goal\}}

\par
\textbf{Setup}

\par
To accomplish this task, you are given a set of \textit{\{self.os\_type.value\}} audit system logs below.

\par
\textbf{Input log format}

\par
\textit{\{input\_format\}}

\par
\textbf{Instructions for reporting \textit{\{self.task\}} actions}

\par
Report your findings, if any, by using one or more JSON objects following the format specified below.

\par
\textit{\{self.output\_description\}}

\begin{verbatim}
{{ {output_format}{mitre_output_str} }}
\end{verbatim}

If no potential \textit{\{self.task\}} activity is found, return a single empty JSON object.
\textit{\{additional\_sections\_str\}}

\par
\textbf{Data}

\par
The input audit log data that should be used for the identification of \textit{\{self.task\}} is as follows: \\
\textit{\{logs\}}

\end{tcolorbox}

\subsection{\prompttwo example for exfiltration task}\label{sec:promptv2_example}
\begin{tcolorbox}[
    breakable,
    width=\linewidth,
    colback=gray!10,
    colframe=gray!50,
    boxrule=0.1pt,
    arc=1mm,
    boxsep=1mm,
    left=1mm,
    right=1mm,
    top=1mm,
    bottom=1mm,
    before skip=1ex,
    after skip=1ex
]
\small
You are a highly-skilled computer security expert with decades of experience in forensic system logs analysis.

\par
\textbf{Goal}

\par
You are tasked with identifying exfiltration activity by a potential attacker, that is, actions that cause user-generated data to leave the system. Look for activities such as external drive shares, USB activity, uploads to websites like Dropbox, iCloud, etc., outgoing data transfer over various protocols like HTTPS, or other types of exfiltration. The user-generated data consist of files and directories that are created by the users of system and the attacker may compress or encrypt them before exfiltration.
 
Do not report the data that are not user-generated, e.g. system-generated files or temporary files.

You must identify all exfiltration actions that may be used by potential attacker to steal the user-generated data, and assess how suspicious each action is, i.e., likely to be caused by an attacker.

\par
\textbf{Setup}

\par
To accomplish this task, you are given a set of linux audit system logs below.

\par
\textbf{Input log format}

\par
The audit log data from a linux system has been processed into an information-flow graph. Each line of input now represents an information flow edge that describes an action between two nodes, and each node corresponds to one of three objects: a process, file and network socket. Each input edge contains the following information:

\par
\texttt{"(timestamp of action) [unique source node ID]\{key-value pairs of source node attributes separated by commas\} --> [unique edge ID]\{key-value pairs of the action's attributes separated by commas\} --> [unique destination node ID]\{key-value pairs of destination node separated by commas\}"}

\par
\textbf{Instructions for reporting exfiltration actions}

\par
Report your findings, if any, by using one or more JSON objects following the format specified below.

\par
For each relevant action in the logs that indicates suspicious data exfiltration behavior, return a JSON object following the format specified below.

\begin{verbatim}
{
    "key log lines": str,  # list of
    comma separated log lines showing 
    exfiltration of all the user-generated
    data involved in the data exfiltration
    activity
    "timestamp": str,  # timestamp of 
    the exfiltration action.
    "pid": int,  # the PID of the process 
    performing the data exfiltration action.
    "exfiltration method": str,  # if the
    user-generated data was  exfiltrated,
    how was it exfiltrated?
    "exfiltrated data": str,  # the 
    full file path of the user-generated
    files that are exfiltrated. To
    generate the full file path, 
    combine the directory path and 
    file name.
    "evidence for": str,  # describe all 
    the evidence from the key log lines 
    that indicates suspicious data 
    exfiltration.
    "evidence against": str,  # describe
    all the evidence from the key
    log lines that indicates either
    not suspicious or not data
    exfiltration.
    "deliberation": str,  
    # analyze the evidence above, 
    taking into account evidence for
    and against, in order to reach 
    a conclusion.
    "verdict": str,  # one of 
    ['LOW_SUSPICIOUS', 'MEDIUM_SUSPICIOUS', 
    'HIGH_SUSPICIOUS']. Use 
    'HIGH_SUSPICIOUS' only when 
    the identified behavior is
    highly likely for an attacker to 
    create malicious exfiltration. 
    Use 'MEDIUM_SUSPICIOUS' when it is
    not clear how feasible it would be 
    for an attacker to abuse. Use 
    'LOW_SUSPICIOUS' when the activity 
    is likely benign, part of regular
    system activity, or generally 
    difficult for the attacker to 
    use as exfiltration.
}
\end{verbatim}
If no potential exfiltration activity is found, return a single empty JSON object.

\par
\textbf{Data}

\par
The input audit log data that should be used for the identification of exfiltration is as follows:\\
\textit{[input data]}
\end{tcolorbox}

\begin{table*}[t]
\centering
\caption{
The effect of log-chunk sizes on Gemini 2.5 Flash and Gemini 2.5 Pro for the raw representation using \prompttwo. (\S~\ref{par:chunk_size}). Across both models, our default chunk size of 1000 lines achieves good performance across the tasks.
}
\label{tab:context-window-raw-combined}
\resizebox{0.8\textwidth}{!}{%
\begin{tabular}{lll|c|c|c|c}
\toprule
\textbf{Task} & \textbf{Model} & \textbf{Metric} & \textbf{500 lines} & \textbf{1000 lines} & \textbf{2000 lines} & \textbf{4000 lines} \\
\midrule
\multirow{2}{*}{Classification} & \multirow{2}{*}{\makecell{Gem 2.5\\Flash}} & TPR & 3/3 (100.00)    & 3/3 (100.00)    & 3/3 (100.00)    & 3/3 (100.00) \\
                                &                                     & FPR & 4/62 (6.45)     & 2/32 (6.25)     & 1/17 (5.88)     & 1/10 (10.00) \\
\midrule
\multirow{2}{*}{L.M.} & \multirow{2}{*}{\makecell{Gem 2.5\\Flash}} & TPR & 1/1 (100.00)    & 0/1 (0.00)      & 1/1 (100.00)    & 1/1 (100.00) \\
                                  &                                     & FPR & 36/30425 (0.12) & 4/30425 (0.01)  & 4/30425 (0.01)  & 4/30425 (0.01) \\
\midrule
\multirow{2}{*}{Persistence} & \multirow{2}{*}{\makecell{Gem 2.5\\Flash}} & TPR & 3/5 (60.00)    & 3/5 (60.00)    & 3/5 (60.00)    & 3/5 (60.00) \\
                             &                                     & FPR & 2/34511 (0.01) & 2/34511 (0.01) & 2/34511 (0.01) & 2/34511 (0.01) \\
\midrule
\multirow{2}{*}{Exfiltration} & \multirow{2}{*}{\makecell{Gem 2.5\\Flash}} & TPR & 2/4 (50.00)    & 4/4 (100.00)   & 2/4 (50.00)    & 2/4 (50.00) \\
                              &                                     & FPR & 15/50267 (0.03) & 5/50267 (0.01) & 7/50267 (0.01) & 9/50267 (0.02) \\
\midrule
\multirow{2}{*}{Classification} & \multirow{2}{*}{\makecell{Gem 2.5\\Pro}} & TPR & 3/3 (100.00)    & 3/3 (100.00)    & 3/3 (100.00)    & 3/3 (100.00) \\
                                &                                   & FPR & 15/62 (24.19)   & 7/32 (21.88)    & 5/17 (29.41)    & 3/10 (30.00) \\
\midrule
\multirow{2}{*}{L.M.} & \multirow{2}{*}{\makecell{Gem 2.5\\Pro}} & TPR & 1/1 (100.00)    & 1/1 (100.00)    & 1/1 (100.00)    & 1/1 (100.00) \\
                                  &                                   & FPR & 2/30425 (0.01)  & 0/30425 (0.00)  & 1/30425 (0.00)  & 0/30425 (0.00) \\
\midrule
\multirow{2}{*}{Persistence} & \multirow{2}{*}{\makecell{Gem 2.5\\Pro}} & TPR & 4/5 (80.00)    & 5/5 (100.00)    & 5/5 (100.00)    & 4/5 (80.00) \\
                             &                                   & FPR & 2/34511 (0.01) & 0/34511 (0.00)  & 3/34511 (0.01)  & 1/34511 (0.00) \\
\midrule
\multirow{2}{*}{Exfiltration} & \multirow{2}{*}{\makecell{Gem 2.5\\Pro}} & TPR & 2/4 (50.00)    & 4/4 (100.00)    & 2/4 (50.00)    & 2/4 (50.00) \\
                              &                                   & FPR & 10/50267 (0.02) & 6/50267 (0.01) & 11/50267 (0.02) & 3/50267 (0.01) \\
\midrule
\bottomrule
\end{tabular}%
}
\end{table*}

\begin{table*}
\centering
\caption{Effects of using a more aggressive (true positive optimized) interpretation of a model's output (\S~\ref{sec:results_prompts}).
The ``$\Delta$ change'' columns shows the absolute change in the metric's value by switching to the more aggressive variant for each model. 
Green text indicates when the aggressive interpretation improves a metric, and red text indicates that it leads to a worse metric value.
Negative values for F1 score and positive values for FPR (higher false positive rates) indicate that using the aggressive interpretation decreases performance.
}
\label{tab:v3prompt_comparison_with_gpt5_lab:full}
\setlength{\tabcolsep}{3pt}
\footnotesize
\begin{tabular}{lll|r|r|r|r|r}
\toprule
Task & Metric & Rep & \thead{\gptmini,\\ $\Delta$ change} & \thead{\geminiflash,\\ $\Delta$ change} & \thead{\llamamav,\\ $\Delta$ change} & \thead{\gpt,\\ $\Delta$ change} & \thead{\geminipro,\\ $\Delta$ change} \\
\midrule
classification & F1 & edge & \textcolor{red}{-0.20} & \textcolor{red}{-0.01} & \textcolor{red}{-0.32} & \textcolor{red}{-0.33} & \textcolor{red}{-0.01} \\
classification & TPR & edge & \textcolor{darkgreen}{30.00} & 0.00 & 0.00 & 0.00 & 0.00 \\
classification & FPR & edge & \textcolor{red}{22.89} & \textcolor{red}{7.83} & \textcolor{red}{66.27} & \textcolor{red}{28.31} & \textcolor{red}{8.43} \\
\hline
\lm & F1 & edge & \textcolor{red}{-0.10} & \textcolor{red}{-0.11} & \textcolor{red}{-0.13} & \textcolor{darkgreen}{0.21} & \textcolor{red}{-0.07} \\
\lm & TPR & edge & \textcolor{darkgreen}{8.33} & 0.00 & \textcolor{darkgreen}{16.67} & \textcolor{darkgreen}{29.17} & 0.00 \\
\lm & FPR & edge & \textcolor{red}{0.06} & \textcolor{red}{0.05} & \textcolor{red}{0.22} & \textcolor{red}{0.02} & \textcolor{red}{0.03} \\
\hline
persistence & F1 & edge & \textcolor{red}{-0.42} & \textcolor{red}{-0.20} & \textcolor{red}{-0.20} & \textcolor{red}{-0.09} & \textcolor{red}{-0.13} \\
persistence & TPR & edge & 0.00 & 0.00 & \textcolor{darkgreen}{11.11} & \textcolor{darkgreen}{11.11} & 0.00 \\
persistence & FPR & edge & \textcolor{red}{0.09} & \textcolor{red}{0.08} & \textcolor{red}{0.18} & \textcolor{red}{0.01} & \textcolor{red}{0.08} \\
\hline
exfiltration & F1 & edge & \textcolor{red}{-0.40} & \textcolor{red}{-0.10} & \textcolor{red}{-0.32} & \textcolor{red}{-0.12} & \textcolor{red}{-0.10} \\
exfiltration & TPR & edge & \textcolor{darkgreen}{28.57} & 0.00 & \textcolor{darkgreen}{42.86} & 0.00 & 0.00 \\
exfiltration & FPR & edge & \textcolor{red}{0.06} & \textcolor{red}{0.01} & \textcolor{red}{0.18} & 0.00 & \textcolor{red}{0.04} \\
\hline
classification & F1 & raw & \textcolor{red}{-0.14} & 0.00 & \textcolor{red}{-0.20} & \textcolor{darkgreen}{0.20} & \textcolor{red}{-0.03} \\
classification & TPR & raw & 0.00 & 0.00 & \textcolor{darkgreen}{33.33} & \textcolor{darkgreen}{33.33} & 0.00 \\
classification & FPR & raw & \textcolor{red}{3.12} & 0.00 & \textcolor{red}{12.50} & 0.00 & \textcolor{red}{3.12} \\
\hline
\lm & F1 & raw & \textcolor{darkgreen}{0.40} & \textcolor{darkgreen}{0.08} & \textcolor{darkgreen}{0.20} & \textcolor{darkgreen}{1.00} & \textcolor{red}{-0.33} \\
\lm & TPR & raw & \textcolor{darkgreen}{100.00} & \textcolor{darkgreen}{100.00} & \textcolor{darkgreen}{100.00} & \textcolor{darkgreen}{100.00} & 0.00 \\
\lm & FPR & raw & \textcolor{red}{0.01} & \textcolor{red}{0.06} & \textcolor{red}{0.03} & 0.00 & 0.00 \\
\hline
persistence & F1 & raw & \textcolor{red}{-0.22} & \textcolor{red}{-0.32} & \textcolor{darkgreen}{0.20} & \textcolor{darkgreen}{0.23} & \textcolor{red}{-0.41} \\
persistence & TPR & raw & \textcolor{darkgreen}{20.00} & \textcolor{darkgreen}{20.00} & \textcolor{darkgreen}{60.00} & \textcolor{darkgreen}{40.00} & 0.00 \\
persistence & FPR & raw & \textcolor{red}{0.03} & \textcolor{red}{0.05} & \textcolor{red}{0.02} & 0.00 & \textcolor{red}{0.02} \\
\hline
exfiltration & F1 & raw & \textcolor{red}{-0.39} & \textcolor{red}{-0.17} & \textcolor{red}{-0.10} & \textcolor{darkgreen}{0.13} & \textcolor{red}{-0.15} \\
exfiltration & TPR & raw & 0.00 & 0.00 & \textcolor{darkgreen}{50.00} & \textcolor{darkgreen}{25.00} & 0.00 \\
exfiltration & FPR & raw & \textcolor{red}{0.04} & \textcolor{red}{0.01} & \textcolor{red}{0.05} & 0.00 & \textcolor{red}{0.01} \\
\bottomrule
\end{tabular}
\end{table*}

\end{document}